\documentclass[preprint]{elsarticle}
\usepackage{graphicx} %  Necessary 18 Apr 2025
\usepackage{amsmath,amssymb}
\usepackage{color}
\usepackage{multirow}
\usepackage{booktabs}
\usepackage{url}
\usepackage{lineno}  % line number 
\usepackage{tipa} % IPA symbols essential for name of GESI/GEDI 
\usepackage{subcaption} % Subcaption
\usepackage{appendix}

\usepackage{ulem} % for strikethrough 

\journal{Speech Communication}

\biboptions{authoryear}

\begin{document}

\begin{frontmatter}

\title{Predicting speech intelligibility in older adults for speech enhancement using the  \\ Gammachirp Envelope Similarity Index, GESI \tnoteref{mytitlenote}}
% \tnotetext[mytitlenote]{Fully documented templates are available in the elsarticle package on \href{http://www.ctan.org/tex-archive/macros/latex/contrib/elsarticle}{CTAN}.}

%%% Group authors per affiliation:
%\author{Elsevier\fnref{myfootnote}}
%\address{Radarweg 29, Amsterdam}
%\fntext[myfootnote]{Since 1880.}

%% or include affiliations in footnotes:
\author[WUm]{Ayako Yamamoto}
\ead{yamamoto.ayako@g.wakayama-u.jp}

\author[WUm]{Fuki Miyazaki}
\ead{miyazaki.fuki@g.wakayama-u.jp}

\author[WUm,WU]{Toshio Irino}
\ead{irino@wakayama-u.ac.jp}

\address[WUm]{Graduate School of Systems Engineering, Wakayama University, Sakaedani 930, Wakayama, Wakayama 640--8510, Japan}

\address[WU]{Faculty of Systems Engineering \& Center for Innovative and Joint Research, Wakayama University, Sakaedani 930, Wakayama, Wakayama 640--8510, Japan}

%%%%%%%%%%%%%%%%%%%%%%%%%%%%%%%%%%
\begin{abstract}
%%%%%%%%%%%%%%%%%%%%%%%%%%%%%%%%%%

We propose an objective intelligibility measure (OIM), called the Gammachirp Envelope Similarity Index (GESI), that can predict speech intelligibility (SI) in older adults. GESI is a bottom-up model based on psychoacoustic knowledge from the peripheral to the central auditory system. It computes the single SI metric using the gammachirp filterbank (GCFB), the modulation filterbank, and the extended cosine similarity measure. It takes into account not only the hearing level represented in the audiogram, but also the temporal processing characteristics captured by the temporal  modulation transfer function (TMTF). 
To evaluate performance, SI experiments were conducted with older adults of various hearing levels using speech-in-noise with ideal speech enhancement on familiarity-controlled Japanese words.
The prediction performance was compared with HASPIw2, which was developed for keyword SI prediction. The results showed that GESI predicted the subjective SI scores more accurately than HASPIw2. GESI was also found to be at least as effective as, if not more effective than, HASPIv2 in predicting English sentence-level SI. 
The effect of introducing TMTF into the GESI algorithm was insignificant, suggesting that TMTF measurements and models are not yet mature. Therefore, it may be necessary to perform TMTF measurements with bandpass noise and to improve the incorporation of temporal characteristics into the model.

\end{abstract}


\end{frontmatter}

%%------ Line number ---------
%% \linenumbers

%%%%%%%%%%%%%%%%%%%%%%%%%%%%%%%%%%
\section{Introduction}
\label{sec:Intro}
%%%%%%%%%%%%%%%%%%%%%%%%%%%%%%%%%%

The number of older adults is increasing in many countries. As a result, the number of people with age-related hearing loss (HL) is expected to increase. HL impairs speech communication, which is the basis of human interaction, and leads to a decline in quality of life (QOL). 
The Lancet Committee \citep{livingston2024dementia} has also reported that HL is one of the major modifiable risk factors for dementia, highlighting the importance of early intervention.
Hearing aids are one of the most important solutions for age-related HL today. However, not everyone with HL uses them.
For example, the usage rate is around 40\% to 50\% in European countries, whereas in Japan it is only 15\%~\citep{japanhear2022japantrack}. Furthermore, satisfaction levels among wearers are lower in Japan, at around 50\%, compared with over 70\% in European countries.
This may be due to differences in policy, medical approach and issues such as cost. Ultimately, however, it may also be because hearing aids do not yet provide sufficient performance to compensate for HL.

To solve this problem, it is now critical to develop the next generation of assistive listening devices that can compensate for the difficulties experienced by people with HL. 
Speech enhancement (SE) and noise reduction algorithms~\citep{loizou2013speech} need to become more robust and effective based on individual hearing characteristics. 
For algorithm evaluation, subjective listening tests to measure speech intelligibility (SI) are essential, but time-consuming and costly.
Therefore, it is important to develop an effective objective intelligibility measure (OIM) that can predict SI for listeners with individually different HL.

Many OIMs have been proposed to evaluate SI when using SE and noise reduction algorithms~\citep{falk2015objective, van2018evaluation}. 
STOI~\citep{taal2011algorithm} and ESTOI~\citep{jensen2016algorithm} have been used for this purpose in many studies. GEDI has also been proposed for more conservative evaluation~\citep{yamamoto2020gedi}.
They are intrusive methods that use both the unprocessed or clean reference signal and the test signal to compute the metric.
They provide a single metric value that can be converted to the SI scores of speech sounds with a monotonic function such as a sigmoid or cumulative Gaussian. Because of this simplicity, many studies compare the proposed and conventional SE algorithms directly using this metric, without converting to an objective SI score, which requires some human SI data.
Although these OIMs work well for predicting SI in normal hearing (NH) listeners, they cannot predict SI in people with varying degrees of HL who truly need good hearing assistive devices.

There are some OIMs that can take the HL into account. \citet{kates2014hearing} proposed HASPI to predict the SI of listeners with HL when using hearing aids.
One reason may be that the prediction by this version of HASPI, which provides two or more independent metrics, requires some subjective SI scores. The SI score is calculated by the weighted sum of these metrics with a sigmoid function, and the weights could only be estimated by subjective SI scores.

More recently, by extending the algorithm in GEDI, a new OIM called GESI (Gammachirp Envelope Similarity Index) has been proposed to provide a single metric suitable for such SE evaluations~\citep{irino2022speech,yamamoto2023gesi}.
These OIMs can reflect the hearing level and the degradation factors of the active mechanism of the cochlea. They are based on an auditory filterbank, some form of temporal modulation analysis, feature extraction, and correlation or similarity between the reference and test signals. The metric is derived solely from bottom-up features extracted by an auditory model, using a small number of explicitly defined and interpretable parameters.

This approach contrasts with the method of training a neural network (NN) using parallel data of sounds and subjective SI scores, as in HASPI version 2~\citep{kates2021hearing}, HASPIw2~\citep{kates2023extending}, and more recent deep neural network (DNN) models (e.g.,~\citealt{huckvale2022elo, tu2022exploiting, kamo2022conformer}). Clarity Prediction Challenge (CPC) ~\citep{barker20221st, barker20242nd} promoted such machine learning OIMs for SI prediction of listeners with HL. The DNN models outperform the simple bottom-up OIMs with the effect of learning algorithms with massive amounts of training data. However, it is well known that the range of good performance is usually limited by the prepared training data, computing power, and memory. In addition, the derived system is a black box, and it is difficult to analyze how the objective SI score is calculated for individual HL listeners. 
Moreover, the OIMs with NNs and DNNs use a large number of uninterpretable parameters, and their values are derived from learning, including higher-order knowledge, such as contextual information within a sentence. Therefore, there may not be a simple monotonic relationship between the improvement due to signal processing and the predicted SI score derived from them.
This may be especially true when overfitting the training data or in untrained conditions. In other words, it may be difficult to interpret the effect of signal processing from the SI scores.

In contrast, OIMs that model bottom-up auditory signal processing provide better insight into how specific dysfunctions affect the SI of listeners with HL. 
With an OIM consisting only of interpretable parameters, the mapping between the degree of performance improvement achieved through signal processing and the predicted SI scores is expected to be simpler and easier to understand than when using NNs and DNNs.
Furthermore, the bottom-up OIMs could be further improved in line with advances in models based on psychophysical and physiological knowledge. They can also be used as part of the DNN methods, which could improve the performance compared to simple end-to-end (i.e., signal-to-SI) DNN methods.

%%%%%% GESI
GESI was developed with this in mind and to measure performance improvements in effective signal processing such as SE and noise suppression.
GESI has been shown to predict subjective SI scores more accurately than STOI, ESTOI, and HASPI ~\citep{irino2022speech,yamamoto2023gesi}. 
However, the evaluation was limited to the SI experiments with NH listeners using simulated HL sounds processed by WHIS~\citep{irino2023hearing}. The variation of the sounds was also limited to simulating the average hearing levels of 70- and 80-year-olds, although there was variability in listening conditions in the laboratory and in remote crowdsourced environments. It remains to be demonstrated whether GESI can predict SI in older adults with individually different HL.

In this study, GESI was extended based on several psychoacoustic findings. SI can be influenced not only by the hearing level that appears on the audiogram, but also by the temporal processing characteristics ~\citep{moore2013introduction}. The extension includes the introduction of the temporal  modulation transfer function (TMTF)~\citep{viemeister1979temporal, morimoto2019twopoint} as a first attempt.
New SI experiments with older adults were conducted to clarify the effects of the SE algorithm on SI.
The SE was performed with the Ideal Ratio Mask (IRM), which was proposed to provide oracle data to train the DNN-based SE algorithms~\citep{wang2014training}. The results can provide information about the upper performance limit of SE algorithms. The main research question is whether GESI can predict the derived subjective SI scores better than HASPIw2~\citep{kates2023extending}, which is the latest version of the OIM that can incorporate the HL.
Furthermore, we would like to discuss GESI's ability to predict sentence-level SI by comparing it with HASPIv2.

%%%%%%%%%%%%%%%%%%%%%%%%%%%%%%%%%%%%%%%%%%%
%%%%% ----------------------------------%%%%%
\section{Proposed and conventional OIMs}
\label{sec:OIM}
%%%%% ----------------------------------%%%%%
We describe the algorithms of GESI \citep{irino2022speech,yamamoto2023gesi} and its extension in sections \ref{sec:GESIbackground}, \ref{sec:GESIflow}, and \ref{sec:GESIalgorithm}, and the competing conventional OIM, HASPIw2 \citep{kates2023extending} in section \ref{sec:HASPIw2}.

%------------------------------------------------------
\subsection{Background of GESI}
\label{sec:GESIbackground}
%------------------------------------------------------

GESI is designed to objectively predict SI in older adults with age-related HL when listening to sounds enhanced with nonlinear SE algorithms. It is an intrusive measure that requires both the test signal and the original reference signal as inputs. GESI is based on the same auditory model as GEDI \citep{yamamoto2020gedi}, which was applicable to NH listeners. While GEDI measures the spectral distortion between the reference and test signals, GESI assesses the similarity of auditory representations that reflect HL characteristics.
In this study, we extended GESI based on several psychoacoustic observations, such as hearing level, the effect of room acoustics, and temporal processing characteristics.
In the following, we first describe the processing flow of the GESI algorithm in section \ref{sec:GESIflow}.
A detailed explanation of the signal processing in each block and its rationale is then provided in section \ref{sec:GESIalgorithm}.

%%--------------------------------------
\begin{figure}[t] 
    \centering
     \includegraphics[scale= 0.7]{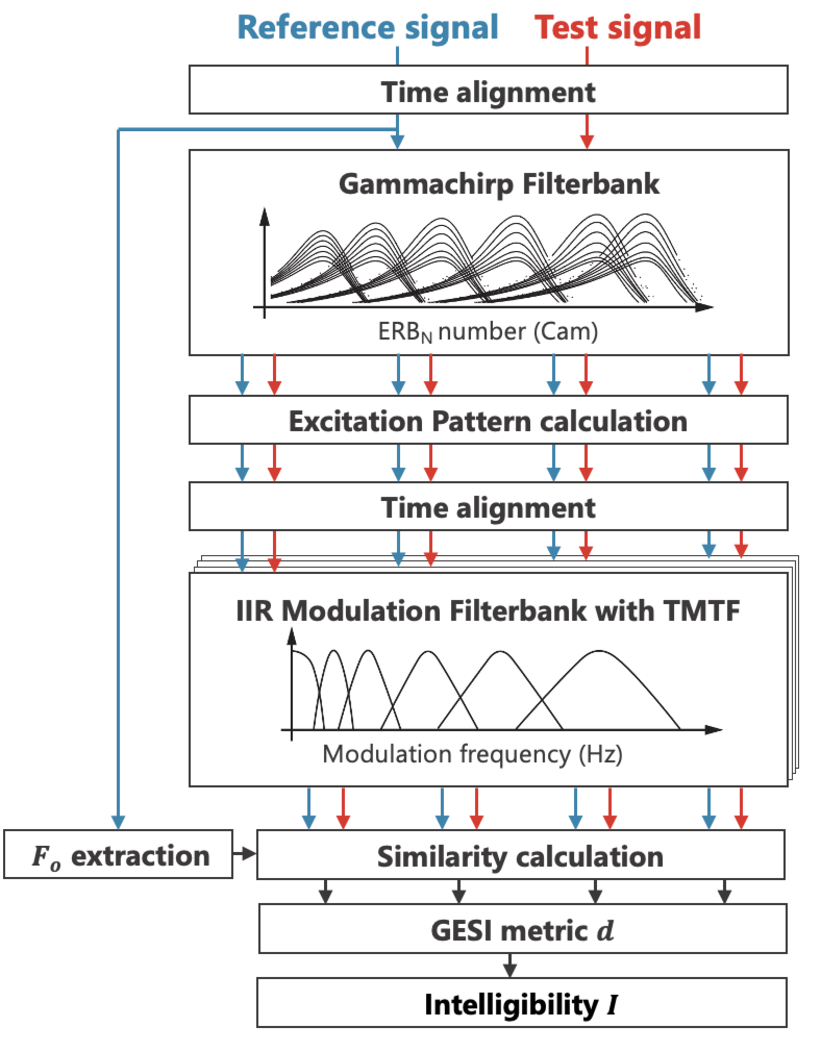}   
     \caption{Block diagram of GESI}
    \label{fig:GESI_BlockDiagram}
\end{figure}
%%--------------------------------------

%------------------------------------------------------
\subsection{Processing flow of GESI}
\label{sec:GESIflow}
%------------------------------------------------------

Figure \ref{fig:GESI_BlockDiagram} shows the block diagram of GESI.
The input sounds to GESI are reference ($r$) and test ($t$) signals.
First, the cross-correlation between them is computed to perform the time alignment of the speech segment. Then, both signals are analyzed with the gammachirp auditory filterbank (GCFB) \citep{irino2006dynamic,irino2023hearing}, which contains the transfer function between the sound field and the cochlea. 
The output is the Excitation Pattern (EP) sequence with 0.5~ms frame shift, something like an ``auditory'' spectrogram (hereafter referred to as EPgram).
The EP is calculated using the cochlear input-output (IO) function with the absolute threshold (AT) set to 0 dB. The noise floor is represented by a Gaussian noise with a root mean squared (RMS) value of 1 for practical calculations.

The reference speech is always analyzed using the GCFB parameters of a typical NH listener, while the test speech is analyzed using parameters that reflect the individual listener's hearing level, that is either normal or with HL. 
This allows GESI to reflect cochlear HL resulting from dysfunction of active amplification by outer hair cells (OHCs) and passive transduction by inner hair cells (IHCs).
The required input parameters are the hearing levels represented by an audiogram, and the compression health parameter ($\alpha$), which indicates the degree of health in the compressive IO function of the cochlea~\citep{irino2023hearing}. 
No dysfunction corresponds to $\alpha=1$ and completely damaged function corresponds to $\alpha=0$. 
In this study, the hearing level was set to 0~dB and $\alpha=1$ (i.e., NH level) to analyze the reference signal. The individual listener's hearing level and a default value of $\alpha=0.5$, a moderate level, were used to analyze the test signal. This is because the $\alpha$ value cannot be estimated without extensive psychoacoustic experiments. See \ref{sec:Estimation_alpha} for more details.

Next, the time offset between the EPgrams of the reference and test speech is compensated for each channel of the GCFB. The cross-correlation is computed to find the peak position within the maximum correction range of $\pm T_{ma}$, and the time alignment is performed accordingly (see section \ref{sec:GESIalgorithm_TimeAlign} for details).

After this correction, the EPgrams are analyzed using an infinite impulse response (IIR) version of the modulation frequency filterbank (MFB) used in GEDI ~\citep{yamamoto2018multiresolution,yamamoto2020gedi}. 
The MFB was first introduced in sEPSM~\citep{jorgensen2011predicting, jorgensen2013multi} and implemented using the Fourier transform. 
However, the IIR version is more phase sensitive, which may lead to better prediction.
The upper limit of the modulation frequency of the MFB was limited to 32 Hz as described in section \ref{sec:GESIalgorithm_LimitMFBrange}.
In addition, the peak gain of each filter in the MFB was set to the corresponding value of the TMTF \citep{morimoto2019twopoint} of NH and older adults, as described in section \ref{sec:GESIalgorithm_TMTF}.

The internal index is computed using an extended version of the cosine similarity between the MFB outputs for the reference signal ($m_{ij}^r(\tau)$) and the test signal ($m_{ij}^t(\tau)$):

%-------- Equation --------
\begin{eqnarray}
  S_{ij} &=& \frac{
  \sum_{\tau} w_i(\tau)\cdot m_{ij}^r(\tau)\cdot m_{ij}^t(\tau)} 
  {(\sum_{\tau}  {m_{ij}^r(\tau)}^2)^{\rho} \cdot (\sum_{\tau} {m_{ij}^t(\tau)}^2)^{\,(1-\rho)}}
  \label{eq:similarity}
\end{eqnarray}
%-------- Equation --------
where $i$ is the GCFB channel $\{i \,|\, 1 \le i \le N \} $, $j$ is the MFB channel $\{j \,|\, 1 \le j \le M \}$, $\tau$ is a time frame number. 
$\rho$ $\{\rho \,|\, 0 \le \rho \le 1\} $ is a weight value that allows us to handle the level difference between the reference and test sounds. Although $\rho=0.5$ is the original definition of cosine similarity, the level difference is normalized to make the evaluation more difficult. See section \ref{sec:GESIalgorithm_Rho} for more details.
$w_i(\tau)$ is a weighting function applied to each GCFB channel, as described in section \ref{sec:GESIalgorithm_SSIweight}. 

The overall similarity index $d$ is obtained by weighting and averaging $S_{ij}$ in Eq.~\ref{eq:similarity} by all $i$ and $j$ .

%-------- Equation --------
\begin{eqnarray}
   d & = & \frac{1}{MN} \sum_{i=1}^N \sum_{j=1}^M  w_j\, S_{ij},
  \label{eq:metric}
\end{eqnarray}
%-------- Equation --------
where $w_j$ is a weighting function applied to each MFB channel.
In this study, $w_j=1$ is used, but is adjustable.

The metric $d$ can be converted to word correct score (\%) or intelligibility $I$ by the sigmoid function used in STOI~\citep{taal2011algorithm} and ESTOI ~\citep{jensen2016algorithm}. 
That is: 
\begin{eqnarray}
   I & = & \frac{I_{max}}{1+\exp(a\cdot d + b)}
   \label{eq:sigmoid}
\end{eqnarray}   
where $a$ and $b$ are parameters determined from a subset of the SI scores in the experimental results using the least squares error method.
$I_{max}$ represents the maximum SI specific to the speech dataset used in the experiments. In this study, it was set to 85\% based on the SI data of the least familiar words in FW07, i.e., a subset of FW03~\citep{amano2009development}.

Conversion to SI is necessary when calculating the RMS error between the subjective and objective SI. However, this is not necessary to compare the performance of SE algorithms. The metric $d$ can be used directly, as is commonly done in much of the literature (e.g., ~\citealp{falk2015objective}), because the SI score $I$ is monotonically related to the metric $d$ as defined in Eq.~\ref{eq:sigmoid}.

%------------------------------------------------------
\subsection{Processing in each block of GESI}
\label{sec:GESIalgorithm}
%------------------------------------------------------
The processing in each block of GESI is explained with extensions from the original version ~\citep{irino2022speech,yamamoto2023gesi}. 

%------------------------------------------------------
\subsubsection{Time alignment of EPgram}
\label{sec:GESIalgorithm_TimeAlign}
%------------------------------------------------------

When listening to speech in a real room environment, reverberation characteristics have a significant impact on speech perception ~\citep{steeneken1980physical}. The room impulse response (RIR) that represents these characteristics varies greatly with location and head orientation and is usually unmeasured and unknown. Therefore, it is necessary to consider the effect in the objective metric.

The EPgrams of the dry reference and reverberant test speech have different time delays across the frequency channels. This causes an initial phase difference at the input of the MFB and results in a degradation of the similarity value $S_{ij}$ in Eq.~\ref{eq:similarity}. To solve this problem, we introduced a time alignment mechanism in each EPgram channel. The preliminary experiments have shown that the prediction accuracy is improved compared to the previous GESI.

This approach is also inspired by the strobed temporal integration mechanism of the Stabilized Wavelet--Mellin Transform (SWMT), a computational model of speech perception \citep{irino2002segregating}. This mechanism synchronizes auditory filterbank outputs in phase. Since the temporal resolution of auditory images in the SWMT is about 30 ms, we set the maximum correction range to $ T_{ma}=\pm30 \rm{ms}$. Without this constraint, similarity calculations could be performed between adjacent different phonemes. This could potentially decrease prediction accuracy.

%------------------------------------------------------
\subsubsection{Upper modulation frequency of the MFB}
\label{sec:GESIalgorithm_LimitMFBrange}
%------------------------------------------------------

It is also important to note that the RIR reduces the modulation depth of speech sounds at the listener's ear. In fact, the Speech Transmission Index (STI) \citep{steeneken1980physical} is a pioneering SI metric that reflects this effect in the equation. In the STI, the upper limit of the modulation frequency is 16 Hz. Based on this, the limit in the current GESI has been set at 32~Hz with a small margin.

%------------------------------------------------------
\subsubsection{Introduction of the TMTF into the MFB}
\label{sec:GESIalgorithm_TMTF}
%------------------------------------------------------

Recently, hidden hearing loss~\citep{liberman2016toward,liberman2020hidden} has been reported in individuals who have difficulty understanding speech in noisy environments, although the decline in audiogram is not as severe. It is often caused by auditory neuropathy or synaptopathy with a decrease in temporal processing characteristics~\citep{zeng1999temporal, narne2013temporal}. 
The SI in older adults with HL may be affected by similar factors, although to a lesser extent.
The TMTF is one of the measures used to capture such characteristics. It has been reported that the modulation detection sensitivity in the TMTF is reduced in older adults with HL compared to NH adults~\citep{morimoto2019twopoint}. Therefore, it seems important to include the TMTF in the OIMs. In this study, we introduced the TMTF of individual older adults into GESI and tested the effect as a first attempt in the history of OIM development.

The TMTF measurement was performed using the two-point method~\citep{morimoto2019twopoint} as described in~\ref{sec:Measure_TMTF}. This method assumes a first-order low-pass filter (LPF) and estimates the modulation depth threshold ($L_{ps}$ in dB) at low modulation frequencies and the LPF cutoff frequency ($F_{c}$ in Hz). 
Because the measurement uses broadband noise, it cannot directly reflect the gain or compression characteristics of individual auditory filters. Therefore, for simplicity, we assumed that the low-pass filter gain of the TMTF is equal to the peak gain of the MFB for all GCFB channels as a first-order approximation.
The peak gain is then formulated for the NH listener to analyze the reference signal ($A_j^r$) and for a HL listener to analyze the test signal ($A_j^t$),

\begin{eqnarray}
    A_j^r & = & \frac{1}{\sqrt{1 + (f_{m_j}/F_{c}^{(NH)})^{2}}} \label{eq:AjRef}\\
    A_j^t & = & \frac{10^{ (L_{ps}^{(NH)}-L_{ps}^{(HL)})/20}} 
        {\sqrt{1 + (f_{m_j}/F_{c}^{(HL)})^{2}}} \label{eq:AjTest}
    \label{eq:A_j_r_t}
\end{eqnarray}
where $j$ is the MFB channel $\{j| 2\le j \le M\}$ and $f_{m_j}$ is the MFB frequency; $L_{ps}^{(NH)}$ and $L_{ps}^{(HL)}$ are the modulation depth thresholds of the NH and HL listeners, respectively, and $F_{c}^{(NH)}$ and $F_{c}^{(HL)}$ are their cutoff frequencies. The peak gain is usually reduced in the test signal analysis since $L_{ps}^{(NH)} < L_{ps}^{(HL)}$ in many cases. Note that the first MFB filter is an LPF with a cutoff frequency of 1~Hz and we set $A_1^r = A_1^t = 1$ to maintain the DC modulation level, which is related to the AT.
%

%------------------------------------------------------
\subsubsection{Weighting function for the GCFB channels}
\label{sec:GESIalgorithm_SSIweight}
%------------------------------------------------------

The weighting function for the GCFB channel $w_i(\tau)$ in Eq.~\ref{eq:similarity} is formulated as the product of two weighting functions, $w_i^{(SSI)}(\tau)$ and $w_i^{(Ef)}$. $w_i^{(SSI)}(\tau)$ is a weighting function designed to reduce the influence of the fundamental frequency $F_o$ (e.g. gender differences) on the phonetic features. $w_i^{(Ef)}$ represents the efficiency of the GCFB channels above the threshold and is described in the next section.

In the aforementioned SWMT, the Size-Shape Image (SSI) was proposed as a representation of speech features \citep{irino2002segregating}. This is a three-dimensional representation in which a two-dimensional image changes over time, allowing an effective representation of speech features independent of $F_o$.
To validate the effectiveness of this representation, an attempt was made to explain psychophysical experimental results related to human size perception \citep{smith2005processing,matsui2022modelling}. 
In the process, a weighting function called ``SSIweight'' was proposed by reducing one dimension of the SSI to fit the two-dimensional representation of EPgram. It was shown that the SSIweight could effectively explain the experimental results regardless of whether the speech was male or female \citep{matsui2022modelling}.

Based on this finding, the previous version of GESI \citep{irino2022speech,yamamoto2023gesi} included the SSIweight, where it was set to a fixed value corresponding to the average $F_o$ of male speech. However, this static approach may not adequately account for within-speech variation or differences in $F_o$ between male and female voices. To address this, the current version introduces $w_i^{(SSI)}(\tau)$, a dynamically varying weighting function defined as a function of frame time $\tau$, formulated as follows.

\begin{eqnarray} 
  w_i'(\tau) & = & \min(\frac{f_{p,i}}{h_{max}\cdot  F_o(\tau)}, 1),\nonumber \\ 
     w_i^{(SSI)}(\tau) & = & \frac{w_i'(\tau)}{\sum_{i=1}^N w_i'(\tau)},
  \label{eq:SSIweight}
\end{eqnarray}

where $f_{p,i}$ represents the peak frequency of 
the $i$-th channel in the GCFB. 
$F_o(\tau)$ denotes the fundamental frequency of the reference speech at frame time $\tau$, which can be estimated using tools such as the WORLD speech synthesizer \citep{morise2016world}. However, for certain sounds, such as some consonants where there is no $F_o$, a small positive value close to zero is assigned.
This ensures that the sum of $w_i^{(SSI)}(\tau)$ for all $i$ is equal to 1.
$h_{max}$ is a constant that determines the boundary between where the weight value $w_i'(\tau)$ gradually increases and where it becomes 1. This comes from the upper limit of the horizontal axis $h$ in the two-dimensional SSI in the SWMT~\citep{irino2002segregating}.

%------------------------------------------------------
\subsubsection{Weighting function for channel efficiency }
\label{sec:GESIalgorithm_AbsThresh}
%------------------------------------------------------
Since the reference speech is analyzed using the GCFB parameters of the NH listener, the EP levels in the EPgrams are always above the AT. In contrast, the test speech is analyzed using the parameters that reflect the individual listener's HL. For older listeners with HL, some high frequency channels may have EP levels falling below the AT, making them inaudible. 

As age-related HL gradually progresses, individuals may compensate by extracting speech information from the remaining audible regions, potentially increasing overall efficiency. We formulate this effect as a weighting function.
Let \( EP_i^{(Ave)} \) be the average EP value over all time frames $\tau$ for the $i$-th GCFB channel (total $N$ channels). Channels where this value exceeds the AT (i.e., 0~dB) can be considered as contributing to the information extraction. Defining the number of such audible channels as \( N_{AT} \), the weighting function \( w_i^{(Ef)} \) is formulated as follows.
\begin{equation}
    w_i^{(Ef)} = 
  \begin{cases}
     (N/N_{AT})^\eta & \text{if $EP_i^{(Ave)} >  AT $ ,} \\
     0                 & \text{otherwise}.
  \end{cases}
  \label{eq:w_i_Ef}
\end{equation}
where $\eta$ is a constant representing the efficiency.
No information can be obtained from channels where $EP_i^{(Ave)}$ is less than or equal to AT, so the weight is set to 0. On the other hand, for channels with values greater than AT, the weight is set as the efficiency-related weight. If $\eta=0$, there is no efficiency improvement and no compensation is made for channel reduction. On the other hand, if $\eta=1$, the efficiency is improved by fully compensating for the channel reduction, making it equivalent to using all channels. However, the audible range is limited and may not reach NH levels.
The efficiency $\eta$ may depend on factors such as listening effort, concentration, and cognitive processes. Therefore, it may be possible to use $\eta$ to introduce cognitive factors into SI prediction using GESI, which is only a bottom-up process.

In this study, $\eta=0.7$ was chosen based on preliminary predictions of subjective SI scores in the current experiment. It was confirmed that $\eta=1$ clearly led to overestimation and $\eta=0$ clearly led to underestimation. Additionally, when $\eta=0.8$ there was not much difference, but it is believed that optimization can be done depending on the experimental participants and conditions.
The effect of variation of  $\eta$ other than 0.7 is discussed in section \ref{sec:PredOIM_Eta}.

%------------------------------------------------------
\subsubsection{Setting of the parameter $\rho$}
\label{sec:GESIalgorithm_Rho}
%------------------------------------------------------

The parameter $\rho$ in Eq.~\ref{eq:similarity} was introduced to correspond to the difference in sound pressure levels between the reference and test signals. 
An important constraint is that the predicted SI score should be low when the test signal level is significantly lower than the reference signal level. If $\rho=0.5$, the levels of reference and test MFB outputs are equalized, and the above constraint cannot be maintained. 
Therefore, setting $\rho$ other than 0.5 is essential.
In the previous study on SI prediction for both laboratory and crowdsourced experiments~\citep{irino2022speech,yamamoto2023gesi}, $\rho$ was defined as a function of tone-pip test results that roughly provide the information about the sensation level (SL) of test sounds for each listener in the given environment.
The experiment in this study was performed in the same quiet laboratory, and the SLs for all listeners were sufficiently above the AT. Therefore, we assumed that accurate predictions could be obtained using a value of $\rho$ similar to that used to predict the SI in the previous laboratory experiment.  Thus, we set the $\rho$ value to 0.55.

%------------------------------------------------------
\subsection{OIM for comparison: HASPIw2}
\label{sec:HASPIw2}
%------------------------------------------------------

The purpose of GESI is to provide information on improvements in SE for older adults with HL. Two important functions are required for such OIMs. First, they must reflect peripheral dysfunctions, including hearing level. Second, they must explicitly represent how the improvement is achieved at the auditory feature level to allow for further improvement. Thus, they must be based on at least a peripheral auditory model. Competitors should have similar functions.

STOI and ESTOI are popular metrics used to evaluate the performance of signal processing. They are simple and effectively used in many SE  studies. However, neither metric accounts for individual hearing levels. Recent DNN-based SI models, such as those proposed in
the first and second CPC (CPC1 and CPC2)~\citep{barker20221st, barker20242nd}, can reflect individual hearing levels. The performance of SI prediction in the competition is excellent. However, it is difficult to understand how the improvements were achieved through signal processing. These models may not provide sufficient information to improve the enhancement process.

HASPI~\citep{kates2014hearing} and its successors are based on an auditory model and may fulfill the above functions.
There are three major versions: the original HASPI~\citep{kates2014hearing}, HASPI version 2 (or HASPIv2) ~\citep{kates2021hearing}, and HASPIw2~\citep{kates2023extending}. 
The original HASPI~\citep{kates2014hearing} is simple and produces several internal metrics that are mapped to an SI score via a sigmoid function. But this function differs from Eq.~\ref{eq:sigmoid} in that it accepts multiple inputs. Because there are no reported preset values for the parameters, it is difficult to use, and stable prediction values cannot be obtained easily. To address this issue, HASPI has been improved and now exists as HASPIv2 and the latest HASPIw2.
Furthermore, previous experiments have shown that the original HASPI has lower predictive performance than the previous version of GESI \citep{irino2022speech}, so it was not considered a competitor.

Both HASPIv2 and HASPIw2 produce a single SI score calculated by using NN although there are 10 independent internal metrics. It is possible to derive the SI score for the current experiment by assuming the NN output is considered as a single internal metric $d$ and transformed into by the sigmoid function in Eq.~\ref{eq:sigmoid}. 
In practice, this method allowed HASPIv2 to serve as the baseline model in CPC2~\citep{barker20242nd}.
Although the internal representations of HASPIv2 and HASPIw2 are difficult to interpret due to the NN in the output stage, they are preferable to DNN models because they are more compact.

The NNs are trained on English speech samples from the HINT database~\citep{nilsson1994development} and the IEEE sentence~\citep{rothauser1969ieee} with different distortions~\citep{kates2021hearing,kates2023extending}.
HASPIv2 is tuned for the SI of the entire sentence, which includes the higher-level linguistic information. In this experiment, the accuracy rate is based on individual words, so consistency cannot be ensured, making it unsuitable.
As the most recent version tuned for keyword SI, HASPIw2 is expected to outperform HASPIv2 in predicting words. Therefore, HASPIw2 was used as the competitor in the SI prediction of the current experiment. 

It would also be important to examine the generalizability of HASPIw2 to other languages and experimental conditions to get an idea of the scope of its application.

%%%%%%%%%%%%%%%%%%%%%%%%%%%%%%%%%%%%
\section{SI experiments with older adults} 
\label{sec:ExpSpIntelEldIRM}
%%%%%%%%%%%%%%%%%%%%%%%%%%%%%%%%%%%%
The SI experiment with older adults was conducted to evaluate the effects of SE using IRM and the performance of OIMs. 
In this experiment, we used unfamiliar Japanese words. When using language, it is impossible to avoid cognitive factors that differ from person to person. However, we wanted to conduct an experiment that controlled for these factors as much as possible.
GESI is a bottom-up model that does not reflect higher-level knowledge. Prediction accuracy is naturally expected to be low when using sentences that incorporate higher-level contextual information. Conversely, monosyllables are not suitable for predicting SI scores in everyday life. Therefore, we selected words that strike this balance. Nevertheless, words that are easily guessed or completely unfamiliar may result in varying SI scores. Thus, we used words from the familiarity-controlled database.

%-----------------------------------------------------
\subsection{Experimental procedure}
\label{sec:Exp_Procedure}
%------------------------------------------------------

%------------------------------------------------------
\subsubsection{Speech materials}
\label{sec:Exp_SpeechMaterial}
%------------------------------------------------------

It is important to minimize cognitive factors such as guessing familiar words in SI experiments used to assess OIMs. There is a Japanese 4 mora word dataset FW07 ~\citep{sakamoto2006new}, for which familiarity is reasonably well controlled. Note that the Japanese mora is a linguistic unit that generally corresponds to a V (vowel) or CV (consonant--vowel) syllable, with some exceptions. The FW07 is a subset of the FW03 dataset ~\citep{amano2009development}, whose properties have been well studied and widely used in Japanese SI experiments. It is also possible to exclude sentence-level context.
We used the speech words pronounced by a male speaker (`mis') from the minimum familiarity rank data in FW07.

To simulate a realistic listening environment, room reverberation was applied to the speech sounds. 
Babble noise was then added to control the SNR. The RIR was obtained from the Aachen database~\citep{jeub2009binaural}. The target speech was convolved with the RIR for 2~m distance in an office room. The babble noise was convolved with the RIRs for 1~m and 3~m separately and added together to have more diffuse characteristics. The signal-to-noise ratio (SNR) was set between -6 dB and 12 dB, increasing in 6 dB steps. The masking effect should be maximized, since the babble noise was created by overlapping and adding a large number of randomly selected word sounds from the FW03 dataset. This condition is referred to as the unprocessed (hereafter ``Unpro'') because no SE algorithm was applied.

We also prepared the word sounds processed by the SE using IRM to evaluate its effectiveness on SI for older adults. The use of IRM has been proposed to provide oracle data for training the DNN-based SE algorithms~\citep{wang2014training}, and its procedure is briefly described in~\ref{sec:IRM}. The ``Unpro'' sounds of the other words were processed with the IRM, and the condition is referred to as ``IRM''.
Since the sound clarity processed by a real SE algorithm should be lower than that of the IRM algorithm, the SI values in the practical situation are expected to be larger than those of ``Unpro'', but smaller than those of ``IRM''.

The IRM-enhanced speech was found to be clearer when processed at a sampling frequency of 16 kHz compared to 48~kHz. Based on this, all of the signal processing described above was performed at 16~kHz and then upsampled to 48~kHz to ensure reliable playback on the experimental website~\citep{yamamoto2021comparison}.

%-----------------------------------------------------
\subsubsection{Data collection}
\label{sec:Exp_DataCollection}
%------------------------------------------------------

Each of the eight conditions, consisting of two signal processing conditions (``Unpro'' and ``IRM'') and four SNR conditions, was assigned a set of 20 words from FW07. Thus, the total number of words was 160 (= 20 $\times$ 2 $\times$ 4), and each participant listened to a different list of words.
The sounds were played on the web-based GUI system ~\citep{yamamoto2021comparison}.
160 words were presented in 16 sessions of 10 words each. 
A word was played and 6 seconds were given to respond, then the next word was played. 
The 6-second response time was longer than the 4-second response time in the previous experiments with NH participants to allow more time to respond. 
Participants listened to the words and wrote them down on a response sheet during this 6-second period.
Even if they did not hear the word completely, they were instructed to guess and respond. After all listening sessions were completed, the answered words were reviewed and entered into the GUI system by the experimenters to avoid unfamiliar keyboard input for the participants. From this data, the correct rates for word, phoneme, and mora, as well as the confusion matrix, were calculated.

Experimental details were explained with documentation, and informed consent was obtained in advance. The experiment was approved by the ethics committee of Wakayama University (Nos.~2015-3, Rei01-01-4J, and Rei02-02-1J).

%-----------------------------------------------------
\subsubsection{Acoustic condition}
\label{sec:Exp_AcousticCondition}
%------------------------------------------------------

The participants were seated in a sound-attenuated room (YAMAHA AVITECS) with a background noise level of approximately 26.2~dB in $L_{\rm Aeq}$.
The sounds were presented diotically through a DA-converter (SONY, Walkman 2018 model NW-A55) connected to a computer (Apple, Mac mini) via headphones (SONY, MDR-1AM2). 
The sound pressure level (SPL) of the unprocessed sounds was 63~dB in ${L_{eq}}$ by default, which was the same level as the calibration tone measured with an artificial ear (Br\"{u}el \& Kj\ae r, Type~4153), a microphone (Br\"{u}el \& Kj\ae r, Type~4192), and a sound level meter (Br\"{u}el \& Kj\ae r, Type~2250-L).
However, four participants reported that the sound was either too loud or too quiet to hear, so the SPL was adjusted to a comfortable level for each participant (62.5, 65, 68, or 70 dB).

% ----------------------------------%
\begin{figure}[t]
%\begin{tabular}{cc}
%  \vspace{5cm}
 \hspace{-1.2cm}
 \begin{minipage}[b]{0.45\linewidth}
 \includegraphics[scale=0.5]{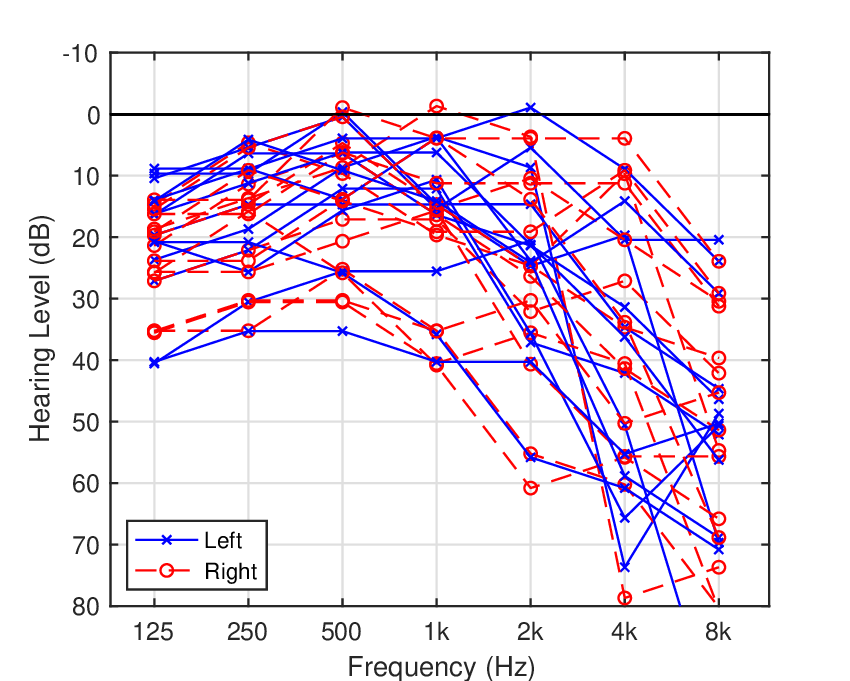}
  \subcaption{Audiograms}
    \label{fig:AudiogramAll}
 \end{minipage} 
% \vspace{-1cm}
  \hspace{1.2cm}
 \begin{minipage}[b]{0.45\linewidth}
 \includegraphics[scale=0.5]{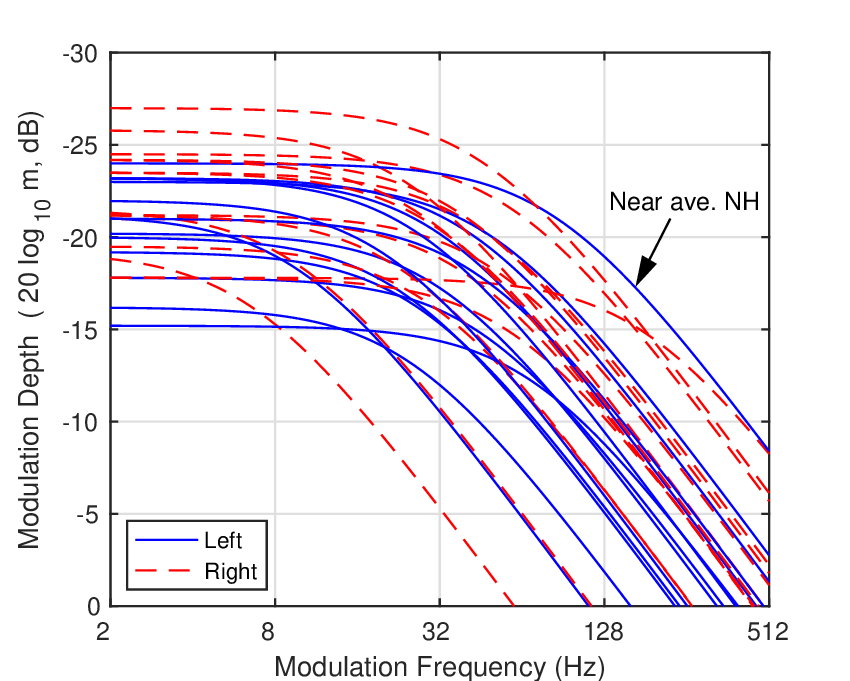}
  \subcaption{TMTF curves }
    \label{fig:TMTFAll}
   \end{minipage}
     \vspace{1cm}
  \caption{Audiograms and TMTF curves for 15 participants for both ears. The blue solid line is for the left ear and the red dashed line is for the right ear.}
\end{figure}
% ----------------------------------%

%------------------------------------------------------
\subsubsection{Participants and hearing characteristics}
\label{sec:Exp_Participant_AudTMTF}
%------------------------------------------------------

The participants in the experiment were 16 people between the ages of 62 and 81 who were recruited through the Senior Human Resources Center in Wakayama. Their native language is Japanese. The audiograms obtained by pure-tone audiometry are shown in Fig.~\ref{fig:AudiogramAll} for 15 participants (one of them was excluded as described shortly).

The hearing levels of the participants varied, but most of them showed typical signs of age-related HL, which is a decrease in the hearing level at 8~kHz.
We calculated the average hearing level from 500 Hz to 4000 Hz for each ear, and the ear with the lower value was considered the better ear. 
Average hearing levels for the better ear ranged from 8.8 to 43.8~dB. Nine participants had levels less than 22~dB, which could be considered NH.

The TMTF was also measured using the two-point method as described in \ref{sec:Measure_TMTF}~\citep{morimoto2019twopoint}. The results are used for SI prediction in GESI (see section~\ref{sec:GESIalgorithm_TMTF}). Note that one participant was unable to perform the TMTF experiment due to difficulty following the procedure and was excluded from the analysis. Therefore, the results reported here are based on 15 participants. 

The TMTF curves for each ear are shown in Fig.~\ref{fig:TMTFAll}. The sensitivity of modulation detection is higher when the curves are at higher positions in the figure. The thresholds of the modulation depth at the low frequency ($L_{ps}$) were in the range from -15 to -27~dB. The cutoff frequencies ($F_c$) also vary widely. The blue line indicated by the arrow is close to the TMTF curve for the average NH listener ($L_{ps}\simeq$ -23~dB and $F_c \simeq$ 128~Hz, see~\ref{sec:Measure_TMTF}).
Because many curves lie below this line, the modulation sensitivity of the older participants is generally lower than that of the NH listener.

% ----------------------------------%
\begin{figure}[p] % page
\begin{minipage}[t]{\hsize}  
% \centering % うまく働かず
\hspace{-1.5cm} % これで位置調節
  \includegraphics[scale=0.6]{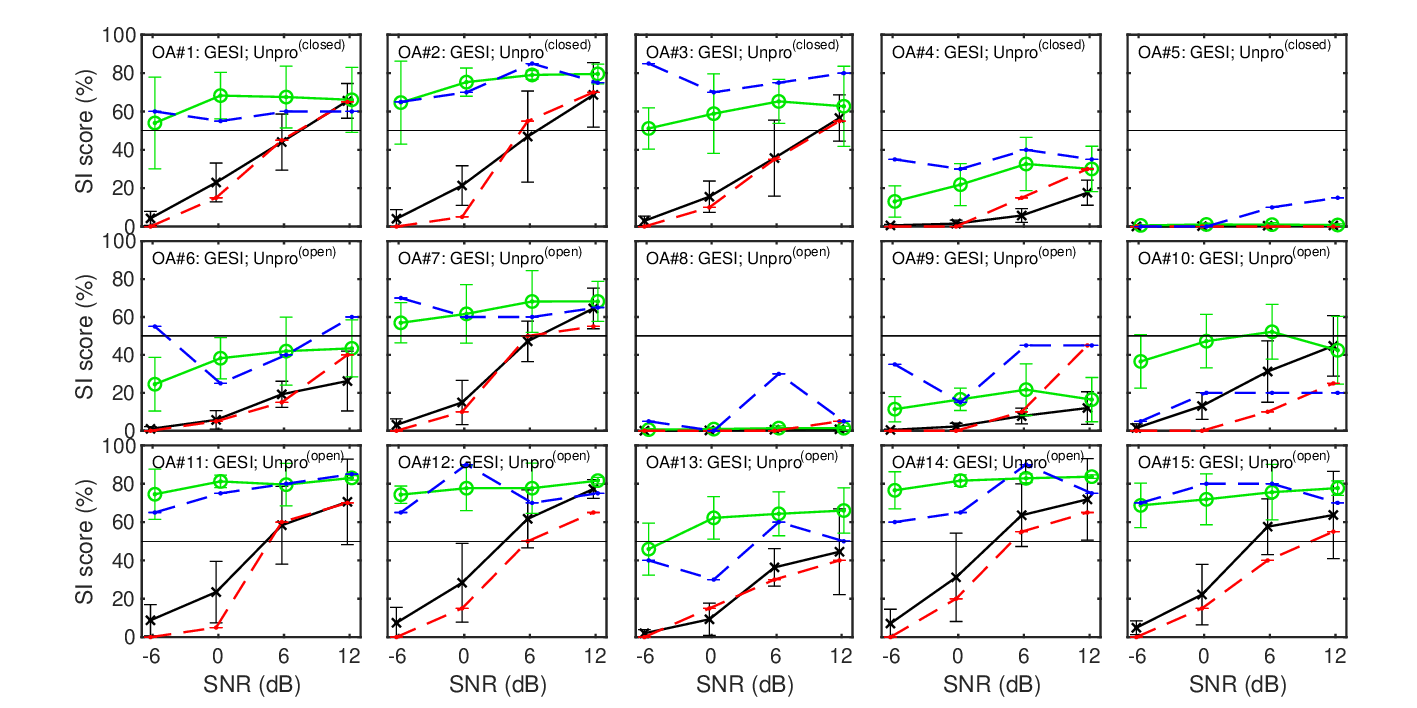}
  \vspace{-15pt}
  \subcaption{Subjective SI and prediction by GESI}
  \label{fig:RsltSbj_GESI}
% \end{figure}

% \begin{figure}[h]
 \hspace{-1.5cm} 
  \includegraphics[scale=0.6]{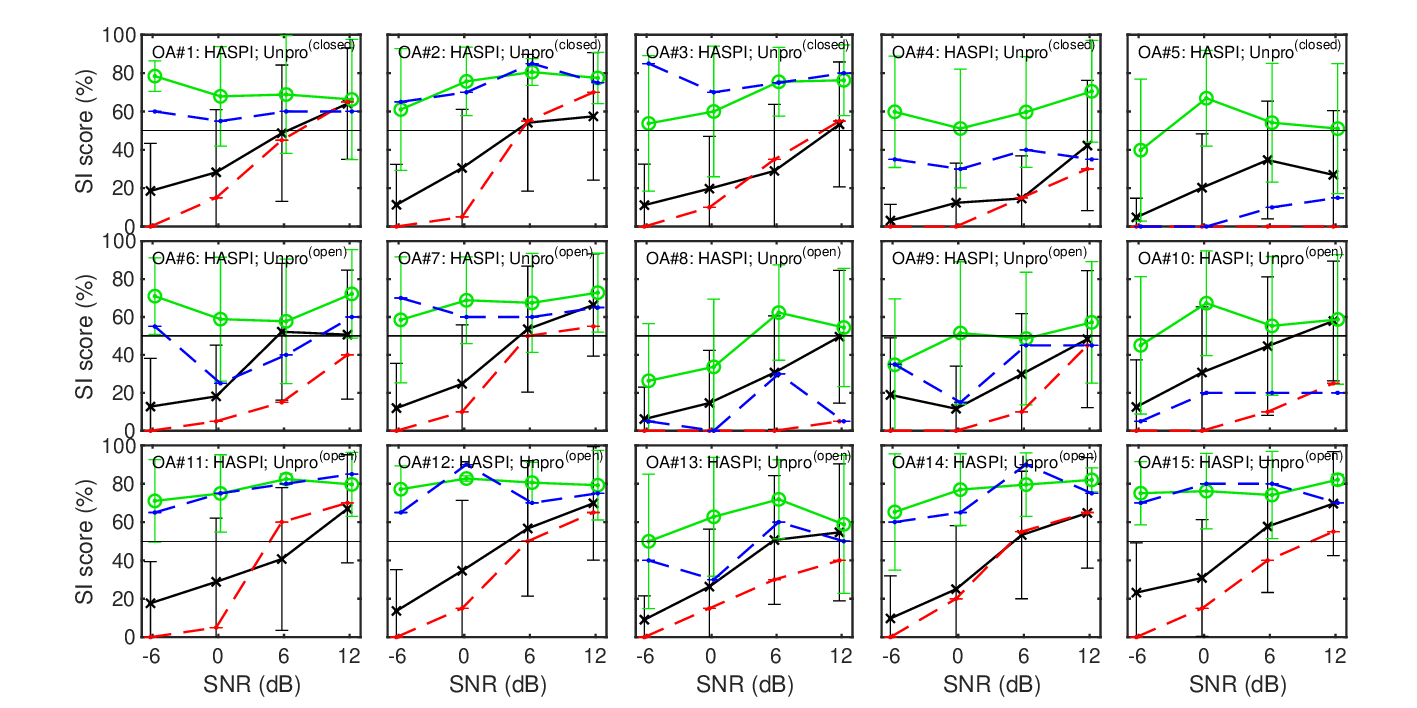}
    \vspace{-15pt}
   \subcaption{Subjective SI and prediction by HASPIw2}
  \label{fig:RsltSbj_HASPIw2}
  \end{minipage}
\caption{SI scores (word correct rate in \%) of 15 participants ($\rm OA^{\#1}$ to $\rm OA^{\#15}$). Subjective experimental results (red dashed line: ``Unpro'', blue dashed line: ``IRM'') and predicted results by OIMs (black solid line with cross: ``Unpro'', green solid line with circle: ``IRM''). Prediction was performed on 20 words for each condition and is presented as mean and standard deviation with error bars.
 }
\label{fig:RsltSbj_OIM}

\end{figure}

% ----------------------------------%

%------------------------------------------------------
\subsection{Results}
\label{sec:ExpResult_EldIRM}
%------------------------------------------------------
We calculated SI as the percentage of 4 mora words that the participants answered correctly. 
Their results are shown as dashed lines in Fig.~\ref{fig:RsltSbj_GESI} and \ref{fig:RsltSbj_HASPIw2} along with the predictions of GESI and HASPIw2 described in the next section.
There are 15 panels each, with participant IDs assigned from $\rm OA^{\#1}$ to $\rm OA^{\#15}$ in order of participation. 
There are large individual differences in SI between participants, which may be due to differences in hearing level, temporal characteristics, and cognitive factors.
The results of the ``Unpro'' condition (red dashed line) show that the SI score improves with increasing SNR. In contrast, the results of the ``IRM'' condition (blue dashed line) show that the SI score does not change much depending on the SNR. This indicates that the effect of the IRM-based SE is more pronounced at lower SNRs. The SI scores for sounds enhanced by a practical SE algorithm are expected to lie between the two SI curves.
These subjective SI scores are the target of the OIM predictions.

%------------------------------------------------------
\section{Prediction by OIMs}
\label{sec:PredictionOIMs}
%------------------------------------------------------

The goal of OIMs is to accurately predict the SI of individual listeners and the effect of SE.
This study investigated whether the subjective SI scores in Figs.~\ref{fig:RsltSbj_GESI} and \ref {fig:RsltSbj_HASPIw2} could be predicted using GESI and HASPIw2. In addition, statistical analysis was performed to confirm the results.

%---------------------------------
\subsection{Condition}
\label{sec:PredOIM_Condition}
%---------------------------------

The hearing levels of the better ear in Fig.~\ref{fig:AudiogramAll} were used for both GESI and HASPIw2. GESI requires compression health ($\alpha$) to specify the slope of the cochlear IO function in GCFB, while HASPIw2 automatically sets the parameter for the similar function. The initial value of $\alpha$ was set to a moderate value of 0.5 for all participants as described in \ref{sec:Estimation_alpha}. 
GESI can also include the individual TMTF shown in Fig.~\ref{fig:TMTFAll}, which has been set to that of the better ear defined in the hearing level.

It is necessary to estimate the sigmoid parameters $a$ and $b$ in Eq.~\ref{eq:sigmoid} in order to convert the internal metric $d$ to the SI score in this experiment. In Figs.~\ref{fig:RsltSbj_GESI} and \ref{fig:RsltSbj_HASPIw2}, the subjective SI scores of the ``Unpro'' condition for the first 5 participants ($\rm OA^{\#1}$ to $\rm OA^{\#5}$) were used for this estimation. 20 words were assigned for each participant and each of the 4 SNR conditions (denoted by ``$\rm Unpro^{(closed)}$''). The $d$ values for these 20 words were calculated and averaged to obtain the average metric $\bar d$. This $\bar d$ was converted to a predicted SI score $\hat I$ using the sigmoid in Eq.~\ref{eq:sigmoid}. The values of $a$ and $b$ were estimated by the least-squares method to minimize the error between $\hat I$ and the subjective SI score $I$ for the five participants and 4 SNR conditions. SI scores for 20 words were then predicted using these $a$ and $b$ values for all participants, SNR conditions, and signal processing conditions (``Unpro'' and ``IRM''). Therefore, the SI scores in ``$\rm Unpro^{(open)}$'' and ``IRM'' were predicted in the open condition.

%---------------------------------
\subsection{Prediction results}
\label{sec:PredOIM_Result}
%---------------------------------
%\color{black}
The prediction results for the 15 participants are presented as mean and standard deviation for 20 words with error bars in Figs.~\ref{fig:RsltSbj_GESI} for GESI and \ref{fig:RsltSbj_HASPIw2} for HASPIw2.
The ``$\rm Unpro^{(closed)}$'' shows the conditions under which the ``Unpro'' SI scores were used to determine the parameters $a$ and $b$ of the sigmoid function in Eq.~\ref{eq:sigmoid}. The ``$\rm Unpro^{(open)}$'' shows the case where the prediction was performed with the open condition using the $a$ and $b$ derived above. Note that the prediction for the ``IRM'' sound was always evaluated with the open condition.

GESI generally predicted quite well for all participants for both ``Unpro'' and ``IRM''. The major differences are only noticeable in the cases of $\rm OA^{\#8}$ and $\rm OA^{\#9}$ for underestimation and $\rm OA^{\#10}$ for overestimation.
In contrast, the predictions of HASPIw2 were not as good as those of GESI. In particular, the standard deviation of 20 words is much larger than that of GESI. The average SI scores for 
$\rm OA^{\#4}$, $\rm OA^{\#5}$, $\rm OA^{\#6}$, $\rm OA^{\#8}$, and $\rm OA^{\#9}$ are much more overestimated than those in GESI. In the case of $\rm OA^{\#5}$, it should have made better predictions even for HASPIw2 because the subjective SI score was used to determine the sigmoid parameters $a$ and $b$ as in ``$\rm Unpro^{(closed)}$'', but it did not. This is probably because the metric value estimated by HASPIw2 is highly variable across words and its mean is always well above zero.
As described in section \ref{sec:HASPIw2}, HASPIw2 uses the NN trained by English keywords. The current results suggest that the generalization ability to this experiment using Japanese words is not very high.

 %---------------------------------
\subsection{Effect of $\eta$ variation}
\label{sec:PredOIM_Eta}
\begin{figure}[t]  % page
\centering
  \includegraphics[scale=0.7]{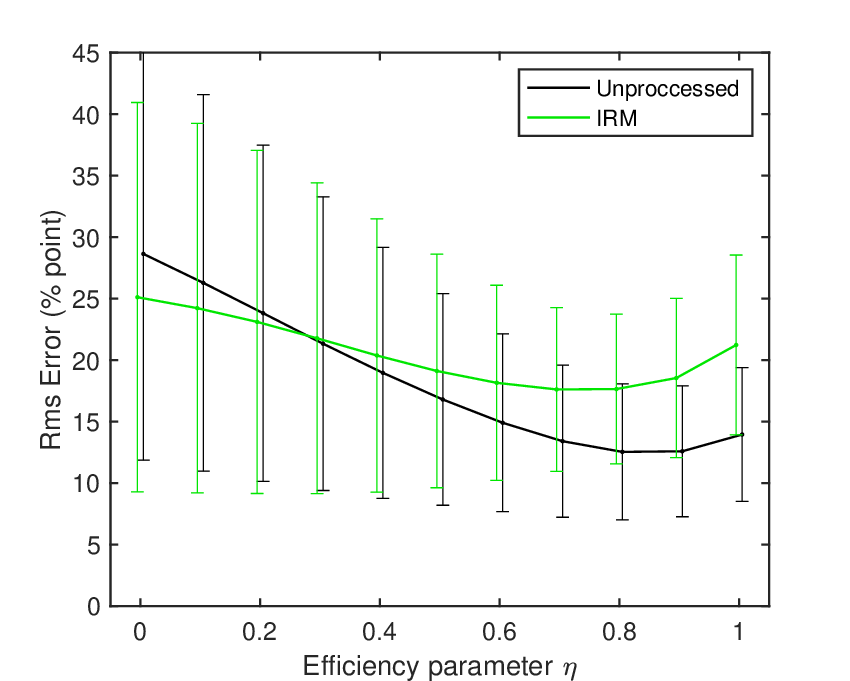}       
 \caption{Root mean square (RMS) error as a function of $\eta$}
\vspace{-5pt}
\label{fig:RMSE_Eta}
\end{figure}

%---------------------------------
The above results were calculated using an efficiency coefficient $\eta$ of 0.7, which was determined in a preliminary study. The validity of this value was not demonstrated. Therefore, we investigated the extent to which $\eta$ affects the error.
Figure \ref{fig:RMSE_Eta} shows the RMS error of individual words when $\eta$ is varied from 0 to 1 in increments of 0.1. The open and closed conditions were the same as in the above prediction shown in Figure \ref{fig:RsltSbj_GESI}.

The curves changed gradually, with no unusual increases or decreases that would suggest overfitting.
In  the ``Unpro'' condition, the minimum value was achieved at $\eta=0.8$, while in the ``IRM'' condition, 
the minimum value was achieved at $\eta=0.7$. 
Since  ``Unpro'' includes both closed and open conditions, whereas ``IRM'' only includes open conditions, its overall RMS error value is lower than that of ``IRM''. The standard deviation is approximately 6\% at $\eta = 0.7$ or $0.8$, which is lower than for the other $\eta$ values. This indicates reduced variability in predictions. Therefore, a value of 0.7 seems to be reasonable for the current prediction.

In this way, GESI's interpretable parameter $\eta$ can be determined through learning using data from a large number of listeners. However, it is thought that adjusting the value of $\eta$ for individual listeners is more effective. This value is thought to allow for the simple introduction of factors that cannot be expressed by the bottom-up model of GESI, such as the influence of the mental lexicon, listening effort, and cognitive factors, into a single value. 
However, this is beyond the scope of this paper, and further research is awaited.

% ----------------------------------%
% ----------------------------------%

% ----------------------------------%
\begin{figure}[p]  % page
\begin{minipage}[t]{\hsize} % for subcaption
\vspace{-2cm}
\centering
  \includegraphics[scale=0.7]{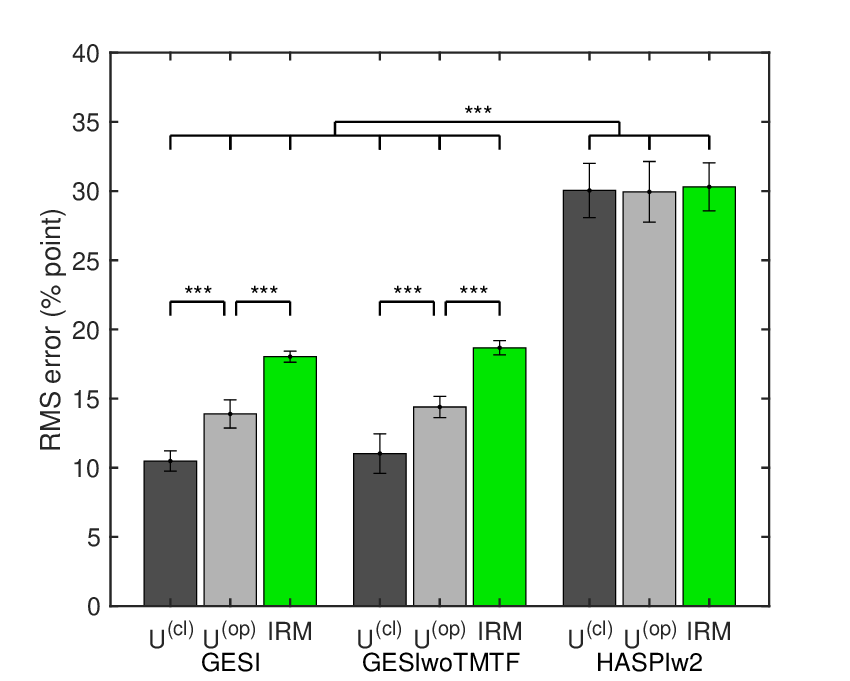}
   \subcaption{RMS error calculated from OIM predictions for individual words }
   \label{fig:RMSE_IndivWord}
%\end{figure}

%\begin{figure}[t]
  \includegraphics[scale=0.7]{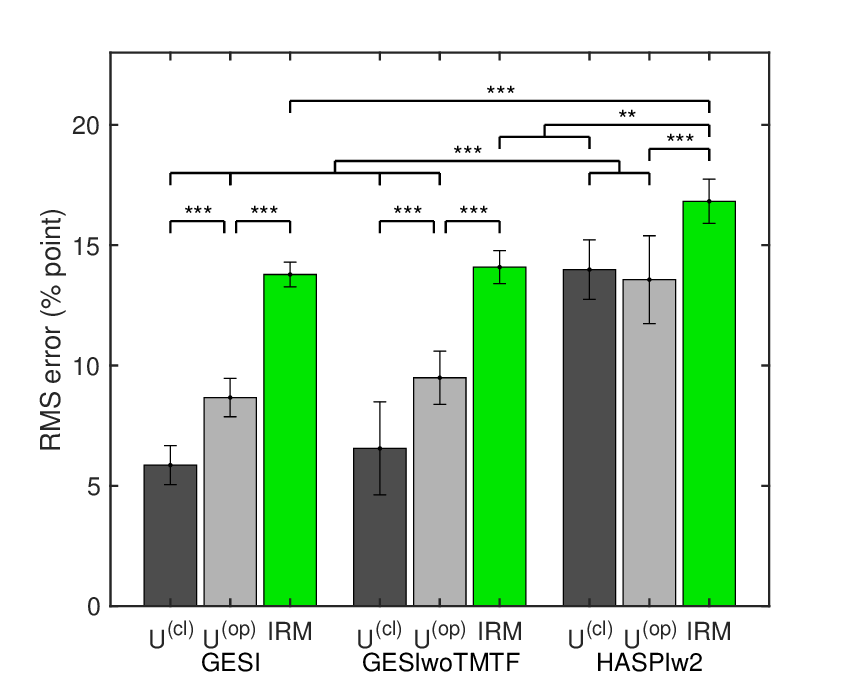}
\subcaption{RMS error calculated from OIM predictions averaged over words in each condition}
   \label{fig:RMSE_MeanWord}
\end{minipage}
          
 \caption{Root mean square (RMS) error between subjective scores and OIM predictions. Bars indicate the mean, and error bars represent the 95\% confidence interval. Results are compared across GESI, GESI without the TMTF, and HASPIw2.
$\rm U^{(cl)}$ and $\rm U^{(op)}$ are the same as ``$\rm Unpro^{(closed)}$'' and ``$\rm Unpro^{(open)}$'' in Fig.~\ref{fig:RsltSbj_OIM}.
Tukey's HSD tests revealed significant differences between GESI and HASPIw2 (**: $p<0.01$, ***: $p<0.001$), but no significant differences between GESI with and without the TMTF under the same process conditions.}
\vspace{-5pt}
\label{fig:RMSE_GESIHASPIw2}
\end{figure}
% ----------------------------------%

%------------------------------------------------------
\subsection{Statistical analysis}
\label{sec:StatAna}
%------------------------------------------------------

Statistical analysis was performed to confirm whether the SI prediction described above is stable and generalizable across different combinations of open and closed sets in GESI and HASPIw2.

For this purpose, different 5 participants were randomly selected to determine the sigmoid parameters $a$ and $b$ from their SI scores of the ``Unpro'' condition, and all SI scores were predicted as described above. This prediction was repeated 9 times for a total of 10 results. 
Furthermore, this procedure was also performed when the TMTF was not introduced in GESI, by substituting the HL parameters for the NH parameters in Eq.~\ref{eq:AjTest}, i.e.~$A_j^t = A_j^r$.
Then the RMS errors were computed in two ways described below and accumulated for three processing conditions of ``$\rm Unpro^{(closed)}$'', ``$\rm Unpro^{(open)}$'' and ``IRM''.

%------------------------------------------------------
\subsubsection{Prediction error of individual word}
\label{sec:StatAna_IndivWord}
%-------------------
The RMS values between the predicted SI scores of individual words and the subjective SI score were calculated for each participant and experimental condition. 
Figure~\ref{fig:RMSE_IndivWord} shows the mean RMS errors and 95\% confidence intervals. 
The mean RMS errors of ``$\rm Unpro^{(closed)}$'', ``$\rm Unpro^{(open)}$'', and ``IRM'' are 10.5, 13.9, and 18.0 \% points in GESI on the 3 leftmost bars. This prediction in ``$\rm Unpro^{(open)}$'' is more difficult than in ``$\rm Unpro^{(closed)}$'', as expected. The prediction in ``IRM'' is the most difficult. However, it is not easy to determine whether an error of 18\% points is so large that it cannot be used practically, especially considering that the standard deviation of the subjective SI scores in the IRM condition was also quite large. The mean RMS errors in HASPIw2 are about 30\% points regardless of the processing condition and much larger than the maximum in GESI. The results reflect that the standard deviations of the predicted SI scores are much larger in HASPIw2 than in GESI, as shown in Fig.~\ref{fig:RsltSbj_OIM}. 
The mean RMS errors in GESI with the TMTF (3 leftmost bars) and without the TMTF (3 middle bars) are about the same.

A three-way analysis of variance (ANOVA) was performed on the following factors: OIM (GESI with and without the TMTF, and HASPIw2), processing condition, and repetition of prediction. The main effects of the OIM, processing, and repetition were significant ($p \ll 0.001$). The interactions between the factors were also significant  ($p < 0.01$), except for the interaction between processing and repetition ($p = 0.20$).
Tukey's HSD multiple comparison tests revealed significant differences between GESI and HASPIw2 ($p<0.001$) for all processing conditions. 
For GESI, there are significant differences between the three processing conditions. However, there is no significant difference within the same processing condition between with and without the TMTF.

%------------------------------------------------------
\subsubsection{Prediction error in mean across words}
\label{sec:StatAna_MeanWord}
%-------------------

The above evaluation may be too strict for HASPIw2 with the larger standard deviation across words. The case where the average of multiple predictions can be used as a metric was also evaluated to relax the condition. The RMS values between the SI score derived from the metric averaged over all 20 words and the subjective SI score were calculated for each participant and experimental condition. 
Figure~\ref{fig:RMSE_MeanWord} shows the results. 
A three-way ANOVA showed results similar to those in the above section.
The difference between GESI and HASPIw2 is smaller than it appears in Fig.~\ref{fig:RMSE_IndivWord}. However, Tukey's HSD tests revealed significant differences between the two under the same processing conditions ($p<0.001$ and $p<0.01$). The difference is expected to increase as the number of words used for averaging decreases. 
This implies that GESI is still better than HASPIw2 in this evaluation.

\color{black}

%------------------------------------------------------
\section{Discussion}
\label{sec:Discussion}
%------------------------------------------------------

%------------------------------------------------------
\subsection{Limitations and Potential of GESI}
\label{sec:Discussion_LimitPotentialGESI}
%------------------------------------------------------

GESI was developed to evaluate the effectiveness of signal processing techniques, such as SE and noise suppression, for older individuals with HL.
GESI is a bottom-up model configured with only a few explicitly defined parameters, in order to clarify the improved internal representation by signal processing. 
The values of the sigmoid parameters $a$ and $b$ in Eq.~\ref{eq:sigmoid} depend on the evaluation material and conditions. They require a few matched data sets of signal and subjective SI scores.
However, these two parameters are unnecessary when GESI is used for performance comparisons.

This feature also highlights a fundamental limitation of GESI.
The current version of GESI does not model the factors involved in central auditory and cognitive processes. Therefore, it is impossible to differentiate the SI scores of listeners with similar hearing level profiles but different central and cognitive dysfunctions or improvements. Furthermore, only two parameters, $a$ and $b$, connect the bottom-up metric and the SI score. These are insufficient for incorporating all the complex, higher-order information that depends on listeners and situations.
This differs from recent DNN-based models, which can make more accurate predictions with a large amount of training data.

Nevertheless, when parameters $a$ and $b$ are determined using a part of individual's SI scores, it is possible to fairly evaluate the effectiveness of signal processing for that individual, as demonstrated under the IRM conditions shown in Fig.~\ref{fig:RsltSbj_OIM}. This may provide valuable information to improve signal processing. This is an important application of GESI.

GESI has another usage.
Depending on the hearing level setting, GESI can predict SI scores for both older adults with HL and young NH individuals. Additionally, GESI only models bottom-up processes. These properties could be used to distinguish between peripheral and central factors contributing to the SI score~\citep{yamamoto2025speech}.
First, the parameters $a$ and $b$ in the sigmoid are estimated using the SI scores of multiple young NHs who are expected to have normal cognitive function. Next, the hearing levels of a specific older adult are set and their SI scores are predicted based on these values of $a$ and $b$.  If the predicted SI score is worse than the score obtained in the subjective experiment, it can be postulated that this older adult has dysfunction in their higher-level processing or cognitive factors. Conversely, a higher predicted SI could suggest greater efficiency in using the mental lexicon or in other higher-level processing in response to gradually declining peripheral hearing level. This method could help to identify the cause of HL.

% \Add{[[[1. Line 713: I would use the word 'postulated' rather than 'estimated'.と指摘されているが、このままで良い気がする。Abstractもそうだし。 --- Line number が違っている --- line 657。  --->  修正済み ２８ Oct 2025]]]}

%------------------------------------------------------
\subsection{Introduction of the TMTF}
\label{sec:Discussion_EffectTMTF}
%------------------------------------------------------

In this study, we tried to incorporate the TMTF of individual listeners into GESI.
However, as shown in \ref{fig:RMSE_GESIHASPIw2}, there was no difference between the GESI results with and without the TMTF.
There are two possible reasons for this.

First, the TMTFs of the older participants were not as poor as those of the neuropathy patients,  whose $N_{ps}$ were greater than -10~dB on average~\citep{narne2013temporal}.
As a result, the effect of the TMTF may not have been large enough to be observed.
However, it remains difficult to conclude whether this is a reasonable explanation.

The second reason is a technical issue that should be improved.
The TMTF introduction method described in section~\ref{sec:GESIalgorithm_TMTF} was imprecise. 
The TMTF measurement described in~\ref{sec:Measure_TMTF} was performed with broadband noise to simplify and expedite the process.
Then, the estimated low-pass characteristics were uniformly reflected in the peak gain of the MFB, regardless of the frequencies in the GCFB channels. This may have negatively impacted the results.
The temporal characteristics may differ from channel to channel.
Furthermore, modulation differences can only be detected at low frequencies within the audible range for older adults with age-related HL. Therefore, the measured results may not reflect the temporal characteristics of high frequencies.
To avoid this issue, the TMTFs should be measured at each audiometric frequency.
If a sinusoid is used for simplicity, the response curve will not become a simple low-pass filter because of the sideband components produced by the modulation~\citep{kohlrausch2000influence}.
However, this does not affect the current version of GESI since the minimum modulation depth remains nearly constant below 32 Hz. Alternatively, bandpass noise could also be used for the measurement.

The method of introducing the TMTF into the MFB should also be considered more carefully. In this study, the measured TMTF was applied directly to the peak gain of the MFB. However, compensation may be necessary because the auditory filter has compressive characteristics.
It is important to take these issues seriously and incorporate them into the design of OIMs.

%------------------------------------------------------
\subsection{Generalization ability}
\label{sec:Discussion_Generality}
%------------------------------------------------------

The results shown in section~\ref{sec:PredictionOIMs} suggest that GESI is more advantageous than HASPIw2 in predicting current experimental results. 
Therefore, HASPIw2 is not sufficiently capable of generalizing to Japanese word SI prediction.
It may be possible to train the NNs in HASPIw2 using compatible Japanese word data to improve performance. 
However, with such a strategy, the training would likely need to be tailored for each language, experimental condition, and listening condition.

Conversely, we also examined whether GESI could generalize to English SI prediction. Here, we compared HASPIv2 and GESI using CPC2 data~\citep{barker20242nd}.
The rationale for using CPC2 data and the comparison with HASPIv2, the evaluation method, and the results are presented in detail in \ref{sec:Eval_CPC2}.  It was difficult to conduct a comparison as rigorous as the prediction presented here because of missing information of the SPL in the CPC2 data. However, the evaluation was designed to be fair by aligning the conditions as much as possible.

The results of \ref{sec:Eval_CPC2} show that the RMS errors in the open test were not significantly different between HASPIv2 and GESI.
In theory, GESI is not superior to HASPIv2 for the CPC2 task because GESI cannot represent higher-order linguistic knowledge, whereas HASPIv2 can. Nevertheless, similar performance was achieved, and GESI is reasonably applicable for predicting English SI with word segmentation.

It is also worth considering the complexity of the models by examining the necessary parameters.
Both models use the sigmoid parameters, $a$ and $b$.
HASPIv2 additionally uses NNs with ten input units (plus one constant unit), four hidden units (plus one constant unit), one output unit. This configuration yields a total of more than 44 uninterpretable weight parameters. In contrast, GESI uses only two explicitly defined parameters $\rho$ and $\eta$.  
This suggests that the same level of performance could be achieved with far fewer parameters. Following Occam's razor principle~\citep{mcfadden2021life}, GESI is clearly simpler and more efficient. Furthermore, this result provides valuable insight into how peripheral processing affects SI.

%------------------------------------------------------
\section{Conclusion}
\label{sec:Conclusion}
%------------------------------------------------------

In this study, a new OIM GESI was extended to predict subjective SI scores in older adults. To assess performance, an SI experiment with familiarity-controlled words was conducted with older adults of different HLs. The TMTF was also measured to see if the temporal response characteristics could be reflected in the prediction. The SI prediction of speech-in-noise and its IRM-enhanced sounds was compared with HASPIw2, which was developed for keyword SI prediction. The results showed that GESI was more accurate than HASPIw2 in the current experiment.
GESI is a bottom-up model based on psychoacoustic knowledge from the peripheral to the central auditory system, with explicitly defined parameters. In contrast, HASPIw2 uses such a bottom-up model with NNs that contain many noninterpretable weight parameters.
The results suggested that its ability to generalize to Japanese word-level SI, i.e., outside the training domain, was limited.  On the other hand, GESI was found to predict English sentence-level SI at a level similar to that of HASPIv2 when using simple word segmentation as described in \ref{sec:Eval_CPC2}.
Introducing the TMTF into the GESI algorithm had no significant effect. This suggests that more effective measurement and implementation methods are needed to incorporate temporal response characteristics.
GESI is available from our GitHub repository~\citep{GitHub_amlab}. 

%%%%%%%%%%%%%%%%%%%%%%%%%%%%%%%%%%
\section*{CRediT author contribution statement}
%%%%%%%%%%%%%%%%%%%%%%%%%%%%%%%%%%

{\bf Ayako Yamamoto: } Writing -- review \& editing, Writing -- original draft, Validation, Software, Methodology, Investigation, Formal analysis, Conceptualization. 
{\bf Fuki Miyazaki: }{Investigation, Data curation.}
{\bf Toshio Irino: }{Writing -- review \& editing, Writing -- original draft,
Visualization, Validation, Supervision, Software, Project administration, Methodology, Investigation, Funding acquisition, Formal analysis, Conceptualization}

%%%%%%%%%%%%%%%%%%%%%%%%%%%%%%%%%%
\section*{Declaration of Generative AI and AI-assisted technologies in the writing process}
%%%%%%%%%%%%%%%%%%%%%%%%%%%%%%%%%%
During the preparation of this work the authors used DeepL Write and DeepL Translate in order to improve language and readability. After using this service, the authors reviewed and edited the content as needed and take full responsibility for the content of the publication.

%%%%%%%%%%%%%%%%%%%%%%%%%%%%%%%%%%
\section*{Declaration of competing interest}
%%%%%%%%%%%%%%%%%%%%%%%%%%%%%%%%%%
The authors declare the following financial interests/personal relationships which may be considered as potential competing interests: Reports a relationship with that includes:. Has patent pending to. If there are other
authors, they declare that they have no known competing financial interests or
personal relationships that could have appeared to influence the work reported in this paper.

%%%%%%%%%%%%%%%%%%%%%%%%%%%%%%%%%%
\section*{Acknowledgments}
%%%%%%%%%%%%%%%%%%%%%%%%%%%%%%%%%%
This research was supported by JSPS KAKENHI: Grant Numbers JP21H03468, JP21K19794,and JP24K02961. The author thanks Takashi Morimoto for the TMTF measurement software and Sota Bano for experimental assistance.

%%%%%%%%%%%%%%%%%%%%%%%%%%%%%%%%%%
% Appendix
% 18 Oct 2025
% Referenceの前にAppendixが必要。
% Speech Communの形式。ーーー 問い合わせあって、そのようにするように返信。
%%%%%%%%%%%%%%%%%%%%%%%%%%%%%%%%%%
\appendix

%%%%%%%%%%%%%%%%%%%%%%%%%%%%%%%%%%
\section{Cochlear HL and estimation of $\alpha$}
\label{sec:Estimation_alpha}
%%%%%%%%%%%%%%%%%%%%%%%%%%%%%%%%%%

GESI is based on GCFB, which reflects cochlear HL resulting from the dysfunction of active amplification by OHCs and passive transduction by IHCs. This can be observed through the compressive IO function of the cochlea. The compression health parameter $\alpha$ was introduced to specify this ratio. In this appendix, we first describe the basic modeling of cochlear HL and the function of $\alpha$. Next, we will demonstrate how to estimate $\alpha$ and discuss the associated challenges. This is why we set it to 0.5 for all older participants. See also ~\cite{irino2023hearing}.

% ----------------------------------%
% \begin{figure}[htbp]
\begin{figure}[t]
%\flushleft
\centering
\includegraphics[scale=0.7]{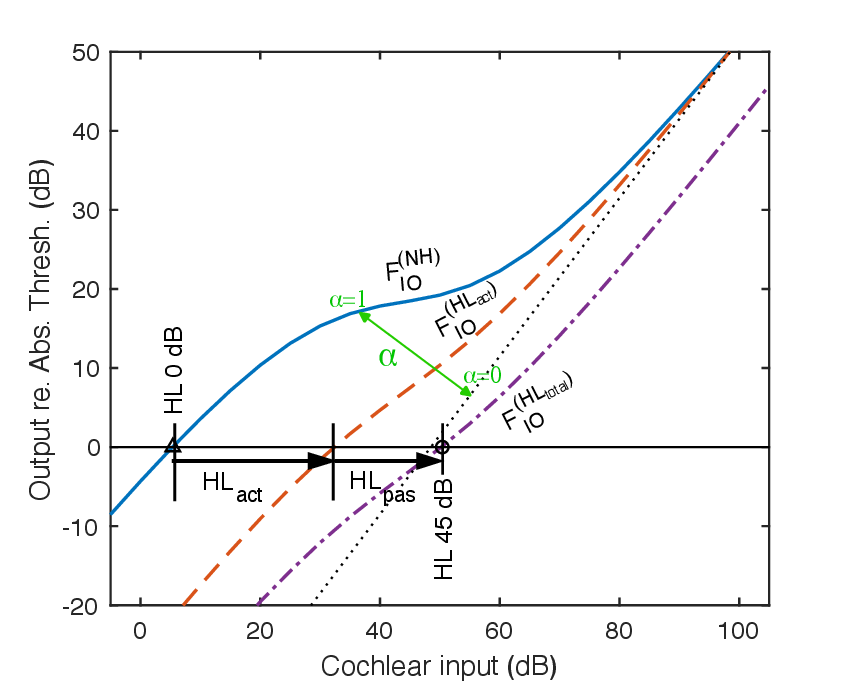}
%\hspace{-120pt}
%\includegraphics[scale=0.7, bb = 0 0 200 200]{FigA1_IOfunc_AsymFunc_HL_ACT_PAS_alpha.eps}
\caption{Schematic plot of cochlear input-output function~\citep{irino2023hearing}}
 \label{fig:IOfunc_schematic}
\end{figure}
% ----------------------------------%

%\hspace{-110pt}
%  \includegraphics[scale=0.7, bb = 0 0 200 200]

%--------------------------------------
\subsection{Cochlear IO function and formulation of HL}
\label{sec:CochlearIO_HL}
%--------------------------------------

Figure \ref{fig:IOfunc_schematic} shows a schematic diagram of the cochlear IO function. The cochlea contains OHCs that amplify the basilar membrane vibrations in response to faint sounds. For an NH person, the IO function ($F_{IO}^{(NH)}$ represented by the blue line) does not increase linearly with increasing input levels; rather, it increases gradually. This is referred to as a compressive characteristic. If there is a dysfunction in the active amplification, the IO function becomes $F_{IO}^{(HL_{act})}$ (red broken line). The compression health parameter $\alpha~\{\alpha | 0 \le \alpha \le 1\}$  (green) was introduced to represent this difference.
When $\alpha=1$, $F_{IO}^{(HL_{act})}$ coincides with $F_{IO}^{(NH)}$, i.e., the NH curve with no dysfunction.  When $\alpha=0$, $F_{IO}^{(HL_{act})}$ coincides with the black dotted line representing linear growth. This indicates that active amplification is completely damaged.

Furthermore, when IHCs, which convert vibration to neural firing, are dysfunctional, the output decreases, resulting in $F_{IO}^{(HL_{total})}$ (purple dotted line). When viewed in terms of the absolute threshold corresponding to the audiogram (on the horizontal line where output is 0 dB), the total hearing loss $HL_{total}$ can be expressed as the sum of the active loss $HL_{act}$ and the passive loss $HL_{pas}$ in dB \citep{irino2023hearing}. This formulation is also used in a loudness model by \cite{moore1997model}.

\begin{equation}
   HL_{total} = HL_{act}+HL_{pas}
    \label{eq:HLtotal_act+pas}
\end{equation}

%--------------------------------------
\subsection{Apportionment of HL}
\label{sec:HL_apportionment}
%--------------------------------------

Although the above formulation is sufficiently simple, it is not possible to determine the ratio of $HL_{act}$ to $HL_{pas}$ or $\alpha$ from an individual audiogram alone.
To determine this, an accurate estimation of the slope of $F_{IO}^{(HL_{act})}$ is necessary. Ideally, the estimation experiments would take as little time as audiometry tests. However, this is challenging as described in the next subsection. Therefore, this study uses the preset value of $\alpha=0.5$, as did previous studies.

Even if the preset is $\alpha=0.5$, the lower limit of $\alpha$ is automatically determined so that it does not fall below the hearing level ($HL_{total}$). Therefore, the practical value of $\alpha$ different among audiometric frequencies and individual listeners (e.g., \citealp{irino2024signal}). 
The proportion of $HL_{act}$ has been reported to be relatively high at low frequencies~\citep{schlittenlacher2020fast}. Therefore, we consider setting $\alpha=0.5$ to be generally adequate, although the exact details remain unclear.

%--------------------------------------
\subsection{Estimation of the IO slope and its limits}
\label{sec:estimationIO}
%--------------------------------------
Three methods for estimating the IO slope have been considered thus far.

\begin{enumerate}

\item Estimating the IO function using the notch noise simultaneous masking method and a level-dependent auditory filter, such as a compressive gammachirp~\citep{patterson2003extending,irino2023improving}.  This method does not involve difficult experimental tasks. However, it requires a large number of measurement points, making it much more time-consuming than an audiometry test.
\item Loudness matching by patients with unilateral HL ~\citep{moore1997modelofloudness}. Since loudness functions and cochlear IO functions are thought to be closely related, this is a reasonable direct measurement method. However, unilateral HL and age-related HL may have different underlying mechanisms. Additionally, age-related HL generally progresses simultaneously in both ears, making this method inapplicable.

\item A method for directly determining the IO function using the time masking curve (TMC) method~\cite{nelson2001new}. This is a forward masking paradigm with a very short detection signal. Estimation results for individuals with moderate HL have also been reported~\cite{lopez2005cochlear}. We implemented this method as well, but the short signals were difficult to detect, even with auxiliary sounds. Considerable training is required for even NH individuals to perform the task, and the method is not simple.
\end{enumerate}

Each method has its own set of challenging drawbacks. However, the IO slope cannot be estimated without conducting such experiments that are much more extensive than SI experiments. A new method is necessary to measure the slope in a time comparable to that of audiometry tests. 

Because these experiments were not conducted as part of this study, it was impossible to estimate the slope or $\alpha$ value for each older adult. However, it could be assumed that older adults may have some degree of active HL. Therefore, it is more reasonable to assume that an $\alpha$ value of 0.5 than to assume they have values of 1 (completely healthy) or 0 (completely damaged).

Note that any recent level-dependent, nonlinear auditory filterbank should consider the slope of the IO function (or equivalently, $\alpha$) to reliably simulate the cochlea.
It is unreasonable to assume that the slope is automatically determined from the hearing level, as HASPIv2 and HASPIw2 do, because these are independent parameters, as described in \ref{sec:HL_apportionment}.

%%%%%%%%%%%%%%%%%%%%%%%%%%%%%%%%%%
\section{Measurement of TMTF}
\label{sec:Measure_TMTF}
%%%%%%%%%%%%%%%%%%%%%%%%%%%%%%%%%%

For estimating the temporal processing characteristics of participants, 
the TMTF measurements~\citep{viemeister1979temporal} were conducted using the two-point method proposed by \cite{morimoto2019twopoint}. 
This method approximates the TMTF as a first-order low-pass filter (LPF) and attempts to determine its parameters using only two measurement points, allowing rapid measurement as in audiometry.

The carrier of the stimulus sound is low noise~\citep{pumplin1985low, kohlrausch1997detection}, which is a broadband noise with minimized envelope fluctuation. The envelope is modulated by a sine wave as 
\begin{eqnarray}
    A(t) & = & A_0 \{1 + m~\sin(2 \pi f_m t)\}, \\
    A_0 & = & \sqrt{1+m^2/2},
    \label{eq:EnvelopeAmpMod}
\end{eqnarray}
where $m$ is a modulation depth, $f_m$ is a modulation frequency, and $t$ is time. $A_0$ is a factor to keep the SPL constant over $m$.

First, the detection threshold of the unmodulated noise was measured for each participant. Using this value, the signal level was set to 20~dB SL so that the signal was presented well above the threshold. Thus, the SPL was individually different. Then, the detection threshold to determine the TMTF was measured by the transformed up--down method with 3-interval, 3-alternative forced-choice and 1~up - 2~down procedure~\citep{levitt1971transformed}.
The modulation depth threshold, $20 \log m$ in dB, was measured in the noise modulated at a frequency of 8 Hz. This value was taken as the peak sensitivity at low modulation frequencies ($L_{ps}$) in dB. Next, the modulation depth was set to $L_{\beta} = L_{ps}/2$ in dB and the modulation frequency was varied to measure the threshold $F_{\beta}$ (Hz), which is the upper limit of the detectable frequency.

From these values, the cutoff frequency of the LPF ($F_c$) can be calculated using the following formula:
\begin{equation}
    F_{c}  = F_{\beta}/\sqrt{10^{-L_{ps}/20} - 1}.
    \label{eq:F_{c}}
\end{equation}
At this point, the TMTF $\varphi(f_{m})$ in dB is given by the following equation:
\begin{equation}
    \varphi(f_{m})  = L_{ps} + 10\log_{10}{\{1 + (f_{m}/F_{c})^{2}\}}
    \label{eq:TMTF}
\end{equation}
Figure~\ref{fig:TMTFAll} shows the TMTF for left and right ears of individual listeners.
Note that the average values for the NH listener were estimated as $L_{ps}\simeq$ -23~dB and $F_c \simeq$ 128~Hz in ~\cite{morimoto2019twopoint}.

%%--------------------------------------------------------------
\section{Speech enhancement using IRM}
\label{sec:IRM}
%%--------------------------------------------------------------

Noisy sounds $x(t)$ can be formulated as 
\begin{equation}
    x(t) = s(t) + n(t) = h(t)*c(t) + n(t)
    \label{eq:xi=si+vi}
\end{equation}
where $t$ is a sample time; $s(t)$, $c(t)$, $h(t)$, and $n(t)$ are the speech sound at the listener's ear, the original clean speech, the impulse response of the acoustic environment, and the noise. The purpose of SE is to reduce the noise $n(t)$ from the observed signal $s(t)$.

A time--frequency representation of this can be derived by applying the short-time Fourier transform.
\begin{equation}
    X_{tf}=S_{tf}+N_{tf}  
\end{equation}
where $t$ and $f$ are the time and frequency frame indexes, respectively.
The Ideal Ratio Mask (IRM)~\citep{wang2014training}  is formulated as:
\begin{equation}
    M_{tf}=\left(\frac{|S_{tf}|^2}{|S_{tf}|^2+|N_{tf}|^2}\right)^{0.5}
    \label{eq:IRM}
\end{equation}
where the powers of the signal $|S_{tf}|^2$ and noise $|N_{tf}|^2$ are known in advance to provide the ideal SE performance.
Using this mask $M_{tf}$, the enhanced time--frequency representation $Y_{tf}$ can be derived as
\begin{equation}
    Y_{tf} = M_{tf} \cdot X_{tf}
\end{equation}
The enhanced sounds are derived by overlapping and adding the inverse short-time Fourier transform of $Y_{tf}$.
In this study, the process was performed at a sampling frequency of 16~kHz with a frame duration of 64 ms (Hanning window) and a frame shift of 16~ms. The sound clarity was better when 16~kHz was used than when 48~kHz was used, probably due to the effect of high frequency components.

%------------------------------------------------------
\section{Evaluation of GESI with CPC2 data}
\label{sec:Eval_CPC2}
%------------------------------------------------------

In the main text, GESI and HASPIw2 were evaluated using familiarity-controlled Japanese words. We found that HASPIw2 does not generalize well to Japanese data. However, it remains unclear whether GESI can generalize sufficiently to English.

%------------------------------------------------------
\subsection{Speech data for evaluation}
\label{sec:Eval_CPC2_speechdata]}
%------------------------------------------------------

The ideal approach to addressing this issue would be to conduct a listening experiment with English words that are controlled for familiarity, similar to the experiment conducted in this study. Although Posoni's group provided a database of familiarity ratings in ~\cite{nusbaum1984sizing}, no speech data were available and no speech-in-noise experiments were performed. More recently, \cite{braza2022effects} conducted a speech-in-noise experiment with familiarity-controlled words. However, their study did not include an evaluation by older adults or the validity of the OIM. Additionally, the speech data used in their experiments is not publicly available.
Given these limitations, conducting a thorough validation study using familiarity-controlled English words would require the development of a new dataset, which is beyond the scope of this study.

Therefore, we chose to use the publicly available CPC2 dataset \citep{barker20242nd} as an alternative. This dataset includes large-scale speech and the corresponding sentence-level (SI) scores of English sentences (see \ref{sec:Eval_CPC2_dataset}). 
However, sentence-level prediction is inherently challenging for bottom-up models like GESI because they do not incorporate linguistic information. We addressed part of this issue with strategies that will be described in \ref{sec:Eval_CPC2_WordWise}.

%------------------------------------------------------
\subsection{Method}
\label{sec:Eval_CPC2_method}
%------------------------------------------------------
We compared HASPIv2 and GESI using the CPC2 dataset.
HASPIv2 contains NNs that were trained using English sentences. It was employed as a baseline model in the CPC2 competition, where performance was evaluated using sentence SI. Although HASPIw2 was used for comparison with GESI in the main text, it is not suitable for predicting sentence SI. Furthermore, there was also a practical issue. CPC2 provides an evaluation environment using Python.
It included a Python version of HASPIv2, but not HASPIw2. Thus, HASPIw2 could not be evaluated. In order to enable a comparison in this environment, we created a Python version of GESI based on the original MATLAB version.

Because of its contextual modeling through NNs, HASPIv2 is expected to be much more advantageous than GESI. Nevertheless, we would like to know how well GESI performs with English sentences.
GESI was originally developed to predict SI for short segments, such as words. Therefore, additional processing was necessary to estimate sentence-level SI. In this study, we employed a simple strategy. First, we segmented each sentence into individual words. Then, we computed GESI scores for each word and combined them to predict the overall sentence SI as described in \ref{sec:Eval_CPC2_prediction_method}.
The following sections provide more detail on the procedures used for GESI and describe the data used in the evaluation.

%------------------------------------------------------
\subsubsection{Word-wise computation and binaural processing}
\label{sec:Eval_CPC2_WordWise}
%------------------------------------------------------

First, we  applied  the Whisper automatic speech recognizer \citep{radford2023whisper} to the reference signal in order to obtain word-level timestamps. Then, we used this information to divide the reference and test signals into word-level segments. The word-level signal pairs were used as input for GESI.
To capture all acoustic features and prevent truncation and clipping at the boundaries, the word duration was extended by 50 ms at the beginning and end.

% binaural
SI scores were calculated independently for the signals from the left and right ears. The higher of the two scores for each word was chosen as the final value. This better-ear approach is also used to compute HASPIv2 scores in the CPC2.

%------------------------------------------------------
\subsubsection{Parameter setting}
\label{sec:Eval_CPC2_parameter}
%------------------------------------------------------

Hearing levels were used in the calculation independently for the left and right ears. Missing data was supplemented by extrapolating or interpolating from existing values. As in the main text and \ref{sec:Estimation_alpha}, a fixed value of $0.5$ was used for the compression health parameter, $\alpha$, for all listeners.
%
% volume SPL
The most confusing aspect of the CPC2 dataset is the lack of information about SPLs at specific digital levels. The metadata of each audio file contained a ``volume'' field ranging from 0 to 100, with a default of 50. However, there is no information about the correspondence between volume and SPL. Therefore, we could not perform SI calculations as precisely as those described in the main text. But for the best guess, we assumed that an RMS value of 1 in a digital signal corresponds to a playback level of 120 dB SPL.

% \Add{in a digital signal}

%------------------------------------------------------
\subsubsection{Ground truth data}
\label{sec:Eval_CPC2_ground_truth}
%------------------------------------------------------

We evaluated the accuracy of the predictions for the subjective response data in the CPC2 challenge.
The CPC2 response data included the following:
(1) the text of the speech presented to listeners (prompt);
(2) the transcribed text of listeners' responses;
(3) the number of words in the prompt (nwords);
(4) the number of correctly identified words (hits);
(5) the percentage of correctly identified words (correctness = hits/nwords*100), ranging from 0 to 100\%.
This correctness is the definition of sentence SI in CPC2 and is used as the ground truth.

%------------------------------------------------------
\subsubsection{Prediction method }
\label{sec:Eval_CPC2_prediction_method}
%------------------------------------------------------

As mentioned earlier, GESI was not designed to predict sentence SI. Therefore, we adapted the following simple strategy: 
First, we calculated the GESI score of each word in the sentence. Then, we set a threshold value, which will be described below. If a word's score is above the threshold, we assume that the word has been correctly identified and count it as a hit. Finally, we calculated the correctness as described above, and treated it as the predicted sentence SI.

This strategy requires determining an adequate threshold.
We compiled a list of all the words in the evaluation dataset. 
For each word, there is a GESI score and a listener's binary ``hit-or-miss'' label. Then, we calculated the ROC curve using the true positive rate (TPR) and the false positive rate (FPR) at various GESI score thresholds.  The optimal threshold was determined to be the one that maximizes the Youden index, ~\citep{youden1950index}, which is equal to the TPR minus the FPR. 
 Note that this strategy does not provide any contextual information within the sentences. Therefore, the GESI prediction did not use that information.

%------------------------------------------------------
\subsection{Dataset}
\label{sec:Eval_CPC2_dataset}
%------------------------------------------------------

First, we briefly describe the original CPC2 dataset~\citep{akeroyd20232nd, barker20242nd} to provide an overview of the speech sounds presented to listeners. Then, we explain how the subset was selected for this study.

%------------------------------------------------------
\subsubsection{Original CPC2 dataset}
\label{sec:Eval_CPC2_dataset_original}
%------------------------------------------------------

The CPC2 dataset includes speech-in-noise data processed by hearing aids, as well as the listening responses for people with a HL. These signals were generated for the 2nd Clarity Enhancement Challenge (CEC2)~\citep{akeroyd20232nd}.
The target speech materials were the Clarity Speech Corpus: seven- to ten-word sentences recorded by British English speakers \citep{graetzer2022dataset}.
These sentences were selected from the British National Corpus XML edition~\citep{BNC2007}. 
The sentences were filtered to exclude those containing one or more unusual words. An unusual word is defined as a word not found in the database of ~\cite{kuvcera1967computational}.
Multiple noise sources, such as music, speech, or domestic appliance noises, were added to these clean utterances at SNRs between -12 dB and +6 dB. The clean utterances were added multiple noise sources, such as music and speech. 
The speech-in-noise signals were convolved with head related impulse responses (HRIRs) to reproduce complex and realistic listening scenarios. 

% \Erase{head and room impulse responses} \Add{head related impulse responses} 

The speech sounds have been processed using ten hearing aid algorithms submitted to CEC2~\citep{akeroyd20232nd}. The algorithms varied depending on whether the approach was single-channel enhancement, multichannel processing, or signal amplification. For the listening tests, personalized stereo signals were produced using pairs of hearing aid input signals and the left and right audiograms of the participants, with HL. Participants took the listening tests in their own homes.
They used headphones and a tablet PC, and did not wear their hearing aids. The listeners controlled the volume on the tablet. This is why the SPL could not be precisely controlled.

The CPC2 dataset included training and evaluation data.
The training data contained listener responses and signal information from the listening tests. However, this information was not included in the evaluation data.

%------------------------------------------------------
\subsubsection{Selected dataset for this evaluation}
\label{sec:Eval_CPC2_dataset_selected}
%------------------------------------------------------

We only use training sets because listener responses are available.
There were three separate training sets, which included responses by fifteen listeners in total. 
We randomly selected ten sentences from each listener, resulting in a subset of 150 sentences. Therefore, the subset contains roughly 1000 words.
This subset was used to compare the sentence-level SI prediction performance of GESI against HASPIv2.

The fifteen listeners were split into two groups. Five listeners were used to determine $a$ and $b$ in the sigmoid function in Eq.~\ref{eq:sigmoid} (i.e., closed evaluation), and the remaining ten listeners were used for open evaluation.

% ----------------------------------%
\begin{figure}[t]
\centering
% \vspace{60pt}
  \includegraphics[scale=0.7]{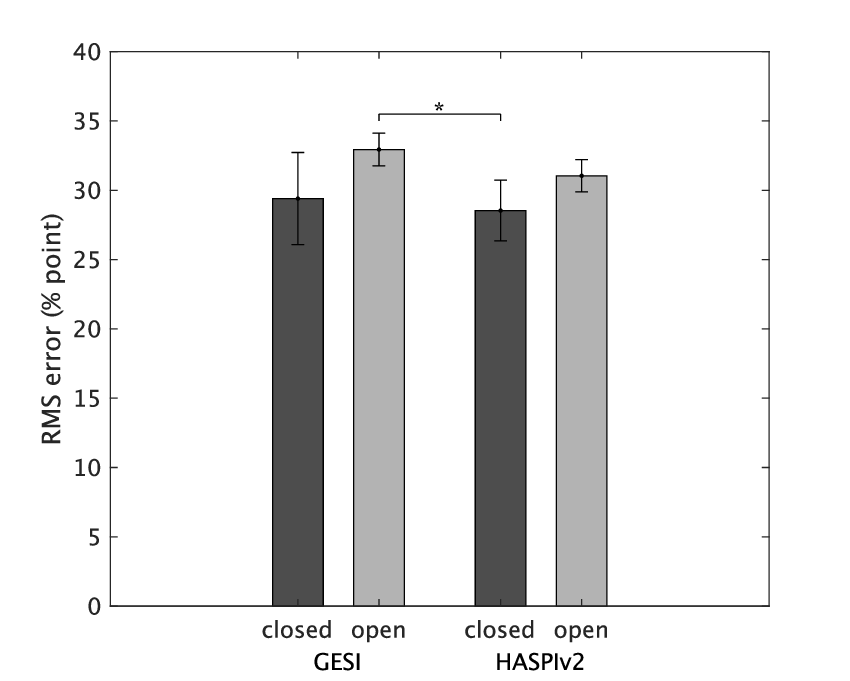}
  \caption{Root mean square (RMS) error between subjective scores and OIM predictions. The bars indicate the mean, and the error bars represent the 95~\% confidence interval. The results were compared across GESI and HASPIv2 for open and closed evaluations. Tukey's HSD tests revealed no significant differences between the conditions, except for one pair (*: $p<0.05$). 
  }
 \label{fig:RMSE_CPC2}
\end{figure}
% ----------------------------------%

%------------------------------------------------------
\subsection{Prediction performance of sentence SI}
\label{sec:Eval_CPC2_result}
%------------------------------------------------------

To ensure a robust evaluation across different data splits, we repeated the random split procedure ten times, as described in  Section~\ref{sec:StatAna}. For each split, we predicted sentence-level SI scores using both GESI and HASPIv2.
We evaluated the prediction performance using the RMS error obtained under each data condition.

For each split, RMS errors were calculated for each listener. Then, the results were averaged. The averaged RMS errors across the ten splits were then used for statistical analysis.
Figure~\ref{fig:RMSE_CPC2} shows the results.
In the closed dataset, the mean RMS error for GESI is 0.87 \% points higher than that for HASPIv2; and in the open dataset, it is 1.9 \% points higher. Since HASPIv2 is trained to predict the sentence-level SI, this result is considered reasonable.
We also conducted a three-way ANOVA to examine the effects of the evaluation condition (closed versus open), the OIM (GESI versus HASPIv2), and repetition of the prediction.
The main effects of the OIM and repetition were not significant. However, the main effect of the evaluation condition was significant ($p < 0.01$).
None of the interactions were significant.
Tukey's HSD tests revealed no significant differences between the conditions, except for one pair, which is shown in Fig.~\ref{fig:RMSE_CPC2}. 
This suggests that GESI is at least as effective as, if not more effective than, HASPIv2.
Considering that GESI did not use contextual information, its performance seems satisfactory.

%%%%%%%%%%%%%%%%%%%%%%%%%%%%%%%%%%
\section*{Data availability}
%%%%%%%%%%%%%%%%%%%%%%%%%%%%%%%%%%
The datasets used and/or analyzed during the current study are available from the corresponding author upon reasonable request. The source of the sound files is the database FW07, provided by the Speech Resource Consortium at the National Institute of Informatics (NII-SRC) under a user license
(DOI: 10.32130/src.FW07). Therefore, you may need a license if you also require stimulus speech sounds.

% 18 Oct 2025
% Referenceの前にAppendixが必要。
% Speech Communの形式。ーーー 問い合わせあって、そのようにするように返信。

%%%%%%%%%%%%%%%%%%%%%%%%%%%%%%%%%%
%\section*{References}
%%%%%%%%%%%%%%%%%%%%%%%%%%%%%%%%%%
%\bibliography{mybibfile}
% \bibliography{Reference_26Mar25} 
%\bibliography{Reference_17Apr25} 
% \bibliography{Reference_21Jul25}
\bibliography{Reference_28Oct25}

\end{document}